\documentstyle[aps,epsfig,prb,amsmath]{revtex}

\input{prmulticol}

\begin{document}
\draft

\title{Chemical bonding, elasticity, and valence force field models: \ a case
study for {\boldmath$\alpha$}-Pt{\boldmath$_2$}Si and PtSi}

\author{J.~E.~Klepeis,$^{(1)}$ O.~Beckstein,$^{(2)}$ O.~Pankratov,$^{(2)}$
and G.~L.~W.~Hart$^{(3)}$}
\address{$^{(1)}$Lawrence Livermore National Laboratory, University of
California, Livermore, CA 94551}
\address{$^{(2)}$University of Erlangen-N\"urnberg, Erlangen, Germany D-91058}
\address{$^{(3)}$National Renewable Energy Laboratory, Golden CO 80401}

\date{10 June 2001}
\maketitle

\begin{abstract}
  We have carried out a detailed study of the chemical bonding for two
  room-temperature stable platinum silicide phases, tetragonal
  $\alpha$-Pt$_2$Si and orthorhombic PtSi.  An analysis of the valence
  electronic charge density reveals surprising evidence of covalent
  three-center bonds in both silicide phases, as well as two-dimensional
  metallic sheets in $\alpha$-Pt$_2$Si.  These elements of the bonding are
  further analyzed by constructing valence force field models using the
  results from recent first principles calculations of the six (nine)
  independent, non-zero elastic constants of $\alpha$-Pt$_2$Si (PtSi).  The
  resulting volume-, radial-, and angular-dependent force constants provide
  insight into the relative strength of various bonding elements as well as
  the trends observed in the elastic constants themselves.  The valence force
  field analysis yields quantitative information about the nature of the
  chemical bonding which is not easily discernable from the more qualitative
  charge density plots.  More generally, this study demonstrates that the
  detailed variations in the elastic constants of a material contain useful
  information about the chemical bonds which can be extracted using valence
  force field models.  Inversely, these models also allow identification of
  specific elements of the chemical bonding with particular trends in the
  elastic constants, both within a given material and among a class of related
  materials.
\end{abstract}
\pacs{PACS 61.50.Lt, 62.20.Dc}

% PACS from http://www.aip.org/pacs/pacs99/pacscheme.html
% 60. CONDENSED MATTER: STRUCTURE, MECHANICAL AND THERMAL PROPERTIES
% 62.   Mechanical and acoustical properties of condensed matter
% 62.20.Dc   Elasticity, elastic constants
%
% 61.50.-f   Crystalline state
% 61.50.Lt   Crystal binding; cohesive energy

% 70. CONDENSED MATTER: ELECTRONIC STRUCTURE, ELECTRICAL, MAGNETIC, AND
%     OPTICAL PROPERTIES
% 71.   Electronic structure
% 71.20.-b   Electron density of states and band structure of crystalline
% solids
% 71.20.Be   Transition metals and alloys
%
% 71.15 Methods of electronic structure calculations
% 71.15.Nc (Total energy and cohesive energy calculations under )

\narrowtext

\section{Introduction}
\label{sec:introduction}

Deposition of metallic platinum silicide compounds on silicon substrates leads
to the formation of rectifying junctions, with a Schottky barrier of
220--240~meV (for holes) in the case of orthorhombic PtSi on p-type
Si~(001).\cite{Pel90,Chi93} \ This energy matches an important atmospheric
``transparency window'' in the infrared region, making these materials well
suited to infrared detector applications.  PtSi has also been discussed as a
promising candidate to replace Ti$_2$Si in polysilicon interconnect
applications in sub-half-micron technologies.\cite{Xu95,Xu96,Mor96} \ In light
of these and other technological applications, as well as a general paucity of
earlier treatments of the fundamental properties of the platinum silicides,
there have been two recent in-depth studies of the atomic and electronic
structure of two room-temperature stable platinum silicide phases,
orthorhombic PtSi and tetragonal $\alpha$-Pt$_2$Si.  Beckstein \emph{et
al.}\cite{Bec01} have carried out an extensive set of first principles
electronic structure calculations for both materials.  In addition to the
electronic structure, they have calculated all of the equilibrium structural
parameters and zero-pressure elastic constants for both phases.  Franco
\emph{et al.}\cite{Fra01} used a combination of photoelectron spectroscopy
(PES), soft x-ray emission spectroscopy (SXE), and x-ray absorption
spectroscopy (XAS) to study the detailed electronic structure of orthorhombic
PtSi.  First principles calculations of the partial density of states (PDOS)
were also carried out in order to aid in interpreting the experimental
spectra.

The present study is complimentary to these two earlier treatments and makes
contact with them in a number of ways.  The combination of the atomic and
electronic structure gives rise to the chemical bonding of a material.  The
elastic constants and the various experimental spectroscopies reflect the
details of this bonding but they do so indirectly.  One of the goals here is
to directly elucidate the fundamental nature of the chemical bonding in the
two silicide phases studied previously.  Towards that end we have calculated
and analyzed the valence electronic charge density for both silicides.
However, this analysis is only qualitative and thus we have made further
attempts to gain a more quantitative understanding.  The previous first
principles study noted a number of interesting trends in the elastic
constants, both within a given material and among the two silicides and the
pure Pt and pure Si phases.\cite{Bec01} \ In the present work we analyze these
trends in much greater detail and in a more quantitative fashion by
constructing valence force field models for all four materials.  The models
are obtained by fitting the first principles elastic constants while also
using insights gained from the charge density analysis to guide the particular
choice of radial and angular interactions.  In turn, the magnitudes of these
various interactions, as obtained from the fits, provide a quantitative
measure of the relative importance of different elements of the chemical
bonding.  In addition, the models can be inverted by expressing the various
elastic constants in terms of the volume-, radial-, and angular-dependent
interactions.  We are thus able to identify the individual trends in the
elastic constants with particular elements of the chemical bonding.

In the present work we have two overall goals.  The first is to gain a
quantitative understanding of the chemical bonding in tetragonal
$\alpha$-Pt$_2$Si and orthorhombic PtSi.  The second goal is to demonstrate,
through a case study of these two silicides as well as pure Pt and pure Si,
that in general terms the variations of the elastic constants of a material
contain useful information about the chemical bonding and that valence force
field models are a convenient means for extracting this information.
Moreover, by inverting the models and identifying the chemical interactions
responsible for the observed trends in the elastic constants we are thereby
able to obtain a more intuitive understanding of the connection between
chemical bonding and the mechanical properties of a material.  Given this more
general goal we have therefore described the construction of the models and
the analysis of various elastic constant trends in some detail.
Sec.~\ref{sec:atomic_structure} provides the relevant details regarding the
atomic structure for the two platinum silicides studied here and in
Sec.~\ref{sec:elastic_constants} we summarize the previous elastic constant
calculations from Ref.~\onlinecite{Bec01}.  The valence electronic charge
densities are analyzed in Sec.~\ref{sec:density} and the valence force field
models are presented in Sec.~\ref{sec:force_field}.  Our results are
summarized in Sec.~\ref{sec:summary}.

\section{Atomic structure}
\label{sec:atomic_structure}

The stable phase of pure Pt at ambient conditions is face-centered cubic
(fcc),\cite{Pea64} while for pure Si it is cubic diamond.\cite{LBnsIII/17a} \
The conventional unit cells of the two platinum silicides $\alpha$-Pt$_{2}$Si
and PtSi are shown in Fig.~\ref{fig:crystal_structures}.  The room-temperature
($T<968$~K) $\alpha$-phase of Pt$_2$Si occurs in the body-centered tetragonal
(bct) structure.\cite{Goh64,Ram78} \ The Strukturbericht designation is
$L'2_{b}$ and the space group is $I4/mmm$ (No.~139).\cite{ITCA} \ The two
symmetry-equivalent Pt atoms in the primitive cell occupy Wyckoff 4(d) sites
and the one Si atom occupies a 2(a) site.  The atom positions are completely
determined by the space group symmetry but there are two independent lattice
constants $a$ and $c$.  PtSi has a primitive orthorhombic structure with four
symmetry-equivalent Pt atoms occupying Wyckoff 4(c) sites and four
symmetry-equivalent Si atoms also occupying 4(c)
sites.\cite{Pfi50,Grae73,Vil91} \ The Strukturbericht designation for this
MnP-type lattice is $B31$ and the space group is $Pnma$ (No.~62).\cite{ITCA} \
The atom coordinates along the ${\mathbf a}$- and ${\mathbf c}$-axes are not
completely specified by the space group symmetry and thus there are four free
internal structural parameters $u_{\text{Pt}}$, $v_{\text{Pt}}$,
$u_{\text{Si}}$, and $v_{\text{Si}}$.  The structure also has three
independent lattice constants $a$, $b$, and $c$.  All of the relevant
equilibrium structural parameters for each of these four materials are given
in Table~\ref{tab:atomic_structure}, including both the experimental values
and the self-consistent theoretical values calculated from first principles in
Ref.~\onlinecite{Bec01}.

\section{Elastic constants}
\label{sec:elastic_constants}

Since we will rely heavily on the detailed results of the first principles
elastic constant calculations from Ref.~\onlinecite{Bec01}, we summarize them
here.  The internal energy $E$ of the crystal is expanded to second order in
the elements of the strain tensor $e_i$, using Voigt notation,
\begin{equation}
  \label{eq:voigt_expansion}
  E(V,\{e_{i}\}) = E(V) + V \sum_i \sigma_{i} e_{i}
                 + \frac{V}{2} \sum_{ij} c_{ij} e_{i} e_{j} + \dots
\end{equation}
where $V$ is the volume of the unstrained crystal, the $\sigma_i$ are the
components of the applied stress tensor, and the $c_{ij}$ are the second order
elastic constants.\cite{Bec01} \ Since the undistorted crystal was always
taken to be the zero-pressure theoretical equilibrium structure, the applied
stress components $\sigma_i$ are all zero and so the second term of
Eq.~(\ref{eq:voigt_expansion}) did not enter.  All of the elastic constant
calculations were carried out using the theoretical equilibrium structural
parameters listed in Table~\ref{tab:atomic_structure}.

Crystals with cubic space group symmetry have only three distinct,
non-vanishing elastic constants.  The theoretical values of these three
elastic constants for both pure Pt and pure Si, as obtained in
Ref.~\onlinecite{Bec01}, are listed in Table~\ref{tab:pt+si_elastic} together
with the corresponding experimental values.  The theoretical bulk moduli were
obtained from the theoretical elastic constants [$B_{0}=\frac{1}{3}(c_{11} + 2
c_{12})$].  We note that in the case of Si the calculation of $c_{44}$
required a relaxation of the positions of the Si atoms within the distorted
unit cell.  The $c_{44}$ strain-induced symmetry reduction allowed the Si
atoms to relax in the $[001]$ direction.  Without this relaxation, $c_{44}$
would have been 108.6~GPa; this is to be compared with the relaxed value of
79.9~GPa and the experimental value of 79.1~GPa.\cite{LBnsIII/29a} \ The
requirement of mechanical stability in a cubic crystal leads to the following
restrictions on the elastic constants\cite{Wal72}
\begin{equation}
  \label{eq:cubic_stability}
  (c_{11} - c_{12}) > 0, \quad c_{11} > 0, \quad c_{44} > 0,
    \quad (c_{11} + 2 c_{12}) > 0.
\end{equation}
These stability conditions also lead to a restriction on the magnitude of
$B_0$ (Ref.~\onlinecite{Bec01})
\begin{equation}
  \label{eq:cubic_b0}
  c_{12} < B_0 < c_{11}.
\end{equation}

Tetragonal $\alpha$-Pt$_2$Si has six independent and non-zero elastic
constants.  Three of these elastic constants, $c_{11}$, $c_{12}$, and
$c_{44}$, correspond to strain-induced symmetry reductions for which the
positions of the Pt atoms are no longer completely fixed by the symmetry.  The
strain-induced forces drive them into energetically more favorable positions
(the corresponding forces on the Si atoms are identically zero by
symmetry). The first principles results for the six elastic constants of
$\alpha$-Pt$_2$Si are given in Table~\ref{tab:silicides_elastic}.  The values
labeled as ``frozen'' correspond to keeping all of the atoms held fixed at the
positions determined solely from the strain tensor, while the elastic
constants labeled ``relaxed'' were obtained by relaxing the strain-induced
forces on the Pt atoms.  The bulk modulus is calculated from the tetragonal
elastic constants, $B_{0}=\frac{1}{9}(2 c_{11} + c_{33} + 2 c_{12} + 4
c_{13})$, and has the same value in the frozen and relaxed calculations.

The requirement that the crystal be stable against any homogeneous elastic
deformation places restrictions on the elastic constants, just as in the cubic
case.  For tetragonal crystals these mechanical stability restrictions are as
follows\cite{Wal72}
\begin{equation}
  \label{eq:pt2si_stability}
  \begin{split}
    (c_{11} - c_{12}) > 0, \quad (c_{11} + c_{33} - 2 c_{13}) &> 0, \\
    c_{11} > 0, \quad c_{33} > 0, \quad c_{44} > 0, \quad c_{66} &> 0, \\
    (2 c_{11} + c_{33} + 2 c_{12} + 4 c_{13}) &> 0.
  \end{split}
\end{equation}
These stability conditions again lead to restrictions on the magnitude of
$B_0$ (Ref.~\onlinecite{Bec01})
\begin{equation}
  \label{eq:pt2si_b0_stability}
  {\frac{1}{3}} (c_{12} + 2 c_{13}) < B_0 < {\frac{1}{3}} (2 c_{11} + c_{33}).
\end{equation}

There are nine independent and non-zero elastic constants for orthorhombic
PtSi.  Relaxation of the internal degrees of freedom was necessary in
calculating all nine PtSi elastic constants because the atomic positions are
not completely fixed by the space group symmetry, even for the unstrained
crystal, and consequently there exist free internal parameters (see
Table~\ref{tab:atomic_structure}) which must be redetermined for any
distortion of the crystal, including hydrostatic pressure.  The results of the
calculations are listed in Table~\ref{tab:silicides_elastic}.  As in the case
of $\alpha$-Pt$_2$Si, the values labeled as ``frozen'' correspond to
calculations in which the internal structural parameters $u_{\text{Pt/Si}}$
and $v_{\text{Pt/Si}}$ were held fixed at their self-consistent equilibrium
values, while in the ``relaxed'' calculations these internal structural
parameters were recalculated to minimize the energy.  The values of $B_0$ are
obtained from the elastic constants, $B_{0}=\frac{1}{9}(c_{11} + c_{22} +
c_{33} + 2 c_{12} + 2 c_{13} + 2 c_{23})$.  As expected, the relaxed value of
$B_0$ is smaller than the frozen value.

Mechanical stability leads to restrictions on the elastic constants, which for
orthorhombic crystals are as follows\cite{Wal72}
\begin{equation}
  \label{eq:ptsi_stability}
  \begin{split}
    (c_{11} + c_{22} - 2 c_{12}) > 0, \quad
    (c_{11} + c_{33} - 2 c_{13}) &> 0, \\
    (c_{22} + c_{33} - 2 c_{23}) &> 0, \\
    c_{11} > 0, \quad c_{22} > 0, \quad c_{33} &> 0, \\
    c_{44} > 0, \quad c_{55} > 0, \quad c_{66} &> 0, \\
    (c_{11} + c_{22} + c_{33} + 2 c_{12} + 2 c_{13} + 2 c_{23}) &> 0.
  \end{split}
\end{equation}
As in the case of $\alpha$-Pt$_2$Si, we can obtain restrictions on the
magnitude of $B_0$,
\begin{equation}
  \label{eq:ptsi_b0_stability}
  {\frac{1}{3}} (c_{12} + c_{13} + c_{23}) < B_0 <
  {\frac{1}{3}} (c_{11} + c_{22} + c_{33}).
\end{equation}

From Tables~\ref{tab:pt+si_elastic} and \ref{tab:silicides_elastic} we see
that overall the elastic constants of $\alpha$-Pt$_2$Si and PtSi appear much
closer to those of pure Pt as compared to pure Si.  The trends of the elastic
constants as a function of the atomic percent Pt in all four materials are
plotted in Fig.~\ref{fig:elas_const_trends}.  Each of the curves corresponds
to an average of a different class of elastic constants, while the symbols
show the values of the individual elastic constants themselves.  As we see
from Eqs.~(\ref{eq:cubic_b0}), (\ref{eq:pt2si_b0_stability}) and
(\ref{eq:ptsi_b0_stability}), mechanical stability requires that $B_0$ be
larger than the average of $c_{11}$, $c_{22}$, and $c_{33}$ but smaller than
the average of $c_{12}$, $c_{13}$, and $c_{23}$ [note that in the case of
$\alpha$-Pt$_2$Si the appropriate averages are ${\frac{1}{3}}(2 c_{11} +
c_{33})$ and ${\frac{1}{3}}(c_{12} + 2 c_{13})$ because $c_{11} = c_{22}$ and
$c_{13} = c_{23}$ for tetragonal crystals].  This stability requirement is
reflected in the top three curves in Fig.~\ref{fig:elas_const_trends}.  We
also see that these three curves each increase monotonically as a function of
atomic percent Pt from pure Si to pure Pt and
we note that all three classes of elastic constants represented by these
curves correspond to strains in which the volume is not fixed.  Conversely,
the two lower curves labeled $(c_{11} - c_{12})/2$ and $c_{44}$ correspond to
the two classes of elastic constants in which the strains are strictly
volume-conserving [in the case of PtSi the lowest solid-line curve and large
open circles correspond to elastic constant combinations ${\frac{1}{4}}(c_{11}
+ c_{22} - 2 c_{12})$, ${\frac{1}{4}}(c_{11} + c_{33} - 2 c_{13})$, and
${\frac{1}{4}}(c_{22} + c_{33} - 2 c_{23})$].  We see that in this case the
two sets of averages are approximately constant as a function of atomic
percent Pt.  The significance of this difference in the trends of
volume-conserving versus non-volume-conserving elastic constants is connected
to the curve labeled $C_0$ and is discussed in Sec.~\ref{sec:force_field}
along with a general discussion of the relationship between the magnitudes of
the various elastic constants and the chemical bonding.

\section{Electronic charge density}
\label{sec:density}

In order to provide insight into the nature of the chemical bonding in
$\alpha$-Pt$_2$Si and PtSi we have analyzed the valence electronic charge
density in these materials.  We have chosen to plot charge density
differences, the superposition of free atom densities subtracted from the
fully self-consistent crystal density, thus emphasizing the formation of
bonds.  Since we are using an all-electron method the calculated charge
density has a large amplitude close to each of the atomic cores.  Subtracting
two such large numbers can sometimes produce unusual features in the plots
described below, but these are of no consequence to our discussion.  Rather we
focus on the smoothly varying density differences in between the atomic cores.
In all of the gray-scale plots the brighter spots represent an increase in the
density relative to superimposed free atoms while the darker spots represent a
decrease, with the exact same scale being used in all of the plots.

\subsection{FPLMTO method}
\label{sec:method}

The valence electronic charge densities were obtained using a full-potential
linear muffin-tin orbital (FPLMTO) method\cite{Met88,Met89} which makes no
shape approximation for the crystal potential.  The crystal is divided up into
regions inside atomic spheres, where Schr\"odinger's equation is solved
numerically, and an interstitial region.  The wavefunctions in the
interstitial are Hankel functions.  An interpolation procedure is used for
evaluating interstitial integrals involving products of Hankel functions.  The
triple-$\kappa$ basis is composed of three sets of $s$, $p$, $d$, and $f$
LMTOs per atom with Hankel function kinetic energies of $-\kappa^2$ = $-$0.01,
$-$1.0, and $-$2.3 Ry (48 orbitals per atom).  The Hankel functions decay
exponentially as $e^{-\kappa r}$.  The angular momentum sums involved in the
interpolation procedure are carried up to a maximum of $\ell$ = 6.  The
calculations presented here are based on the local density approximation
(LDA), using the exchange-correlation potential of Ceperley and
Alder\cite{Cep80} as parameterized by Vosko, Wilk, and Nusair.\cite{Vos80} \
The scalar-relativistic Schr\"odinger equation was solved self-consistently.
We did not include spin-orbit interactions and we used atomic sphere radii
equal to one-half the nearest-neighbor bond lengths.  In the case of
$\alpha$-Pt$_2$Si we included an empty atomic sphere at the octahedral
interstitial site, as well as the usual empty spheres in the interstitial
sites of pure cubic-diamond-phase Si.  However, these empty spheres do not
contribute to the basis but merely improve the accuracy of the interstitial
interpolation procedure.

The Pt 6$s$, 6$p$, 5$d$, and 5$f$ orbitals as well as the Si 3$s$, 3$p$, 3$d$,
and 4$f$ orbitals were all treated as valence states.  The semi-core Pt 5$s$
and 5$p$ orbitals were treated as full band states by carrying out a
``two-panel'' calculation.  The second panel band calculation for the
semi-core orbitals included the Pt 5$s$, 5$p$, 5$d$, and 5$f$ orbitals as well
as all of the Si valence orbitals.  The Brillouin zone (BZ) sums were carried
out using the tetrahedron method.\cite{Bloe94} \ We used the same mesh of
${\mathbf k}$-points for both the self-consistent total energy and charge
density calculations.  In the case of $\alpha$-Pt$_2$Si we used a shifted
24$\times$24$\times$24 (12$\times$12$\times$12) mesh in the full BZ, resulting
in 1056 (159) irreducible ${\mathbf k}$-points in the first (second) panel.
In the PtSi calculations we used a shifted 12$\times$16$\times$12
(6$\times$8$\times$6) mesh in the full BZ, resulting in 288 (36) irreducible
${\mathbf k}$-points in the first (second) panel.  A shifted
28$\times$28$\times$28 (16$\times$16$\times$16) mesh in the full BZ was used
for fcc Pt, resulting in 2030 (408) irreducible ${\mathbf k}$-points in the
first (second) panel.  Finally, a shifted 12$\times$12$\times$12 mesh in the
full BZ was used for cubic-diamond-phase Si, resulting in 182 irreducible
${\mathbf k}$-points.

\subsection{Pt and Si}
\label{sec:pt+si_density}

\begin{figure}[tbp]
  \centerline{\includegraphics[width=\linewidth]{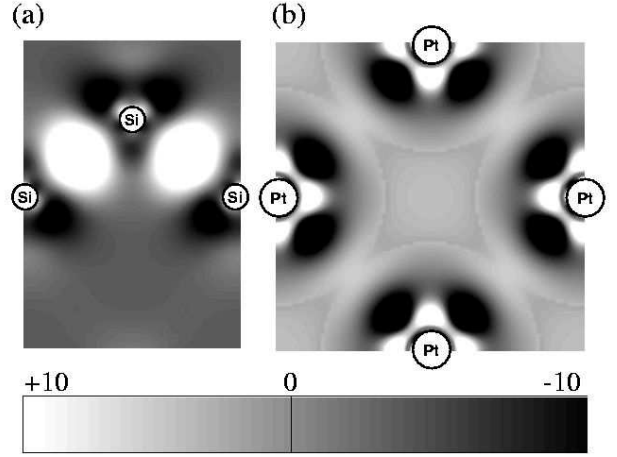}}%
  \caption{Superposition of free atom densities subtracted from the fully
    self-consistent crystal density for (a) cubic diamond-phase Si and (b) fcc
    Pt.  In both plots there are 51 contour levels plotted with pure black
    corresponding to $-10$ and pure white to $+10$ millielectrons/bohr$^3$, as
    indicated in the scale bar.  In (a) the ${\mathbf x}$-axis is along
    $[110]$ and the ${\mathbf y}$-axis along $[001]$ while in (b) the
    ${\mathbf x}$-axis is $[100]$ and the ${\mathbf y}$-axis $[010]$.  In both
    cases the calculations were carried out at the experimental equilibrium
    volume and only the density from the valence states was considered,
    excluding the density arising from the core states.  In the case of Si (a)
    the density was calculated at 76$\times$101 grid points while for Pt (b)
    there were 101$\times$101 grid points.}
  \label{fig:pt+si_density}
\end{figure}

We start with the well-known cases of pure diamond-phase Si and fcc Pt in
order to provide a baseline with which to compare the results we obtain for
the silicides.  In Fig.~\ref{fig:pt+si_density}(a) we see the localized piling
up of additional charge between each pair of Si atoms which corresponds to the
covalent bonds in this material.  Except for these bonds, the density is
relatively unchanged from the free-atom superposition in the remaining regions
outside of the atomic cores, as can be seen by identifying the ``0'' level in
the accompanying scale bar.  This circumstance is in stark contrast to the
case of fcc Pt in Fig.~\ref{fig:pt+si_density}(b).  In Pt the increase in
density is spread approximately uniformly throughout all of the regions
outside the atomic cores.  In fact, from this perspective Pt appears almost
free-electron-like, despite the more localized nature of the states arising
from the partially occupied $d$-band.  Thus we see that charge density
difference plots such as those in Fig.~\ref{fig:pt+si_density} are clearly
able to distinguish metallic bonding, as occurs in Pt, from covalent bonding,
as occurs in Si.  For later purposes we note that the nearest-neighbor spacing
is 2.35~{\AA} in Si and 2.77~{\AA} in Pt.

\subsection{{\boldmath$\alpha$}-Pt{\boldmath$_2$}Si}
\label{sec:pt2si_density}

\begin{figure}[tbp]
  \centerline{\includegraphics[width=\linewidth]{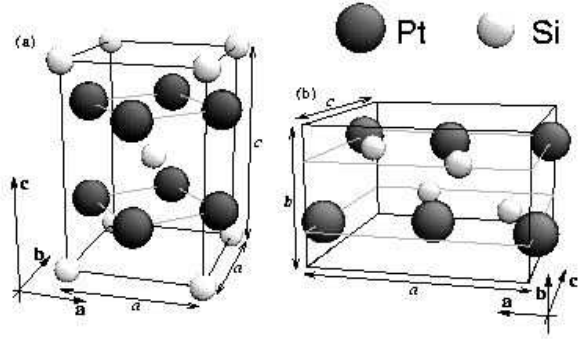}}
  \caption{Conventional unit cells of (a) body-centered tetragonal
    $\alpha$-Pt$_2$Si and (b) orthorhombic PtSi.  The relevant lattice
    constant distances are illustrated in both cases.}
  \label{fig:crystal_structures}
\end{figure}

\begin{figure}[tbp]
  \centerline{\includegraphics[width=\linewidth]{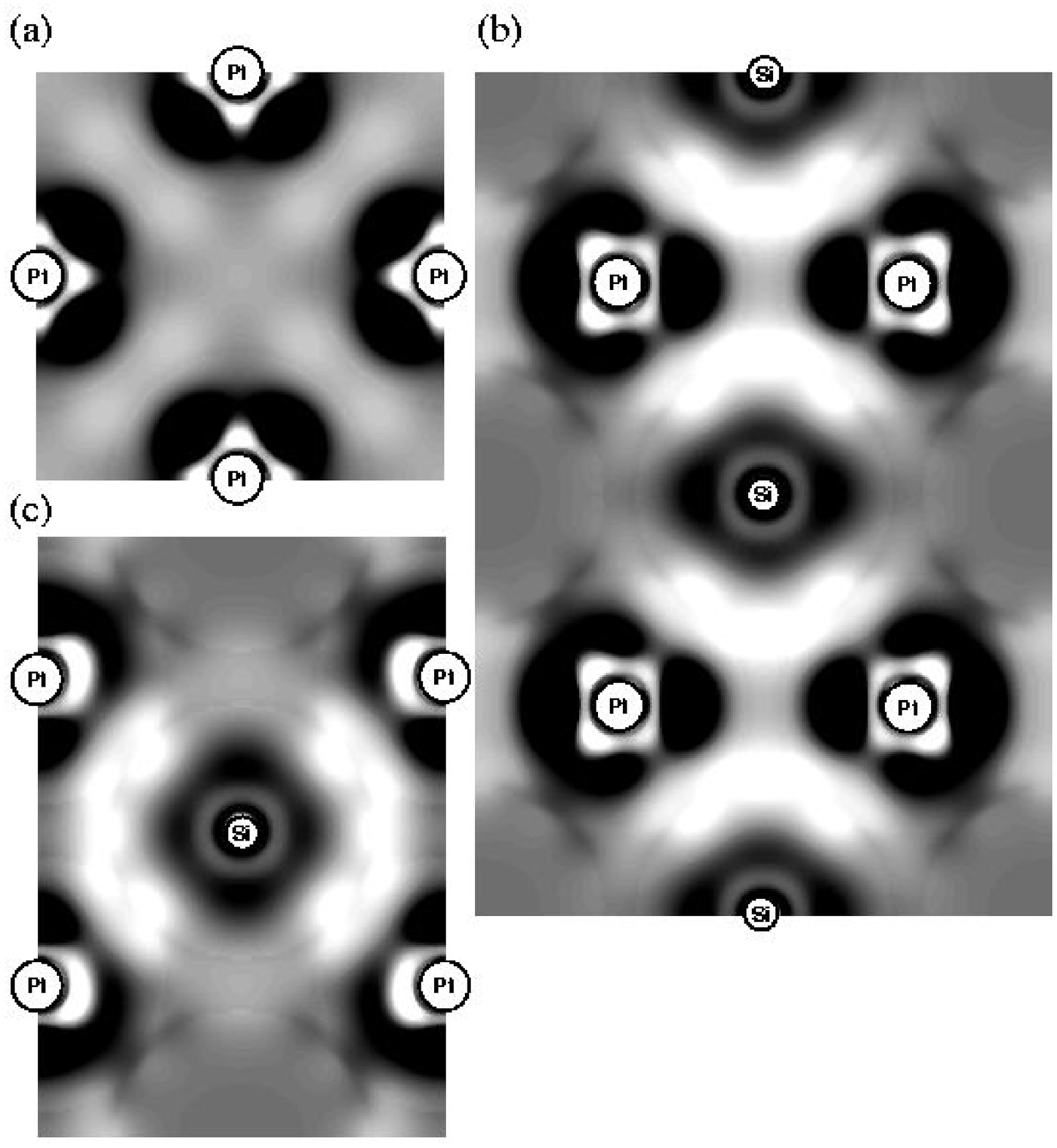}}
  \caption{Superposition of free atom densities subtracted from the fully
    self-consistent crystal density for tetragonal $\alpha$-Pt$_2$Si.  The
    same 51 contour levels and gray-scale are used as in
    Fig.~\protect\ref{fig:pt+si_density}.  The two-dimensional Pt--Pt
    second-nearest-neighbor metallic sheets are shown in (a) with the
    ${\mathbf x}$-axis along $[100]$ and the ${\mathbf y}$-axis along $[010]$.
    In (b) we show the three-center Pt--Si--Pt covalent bonds involving two
    second-neighbor Pt atoms, with the ${\mathbf x}$-axis along $[1\bar{1}0]$
    and the ${\mathbf y}$-axis along $[111]$.  The second set of three-center
    Pt--Si--Pt covalent bonds involving two third-neighbor Pt atoms is
    illustrated in (c) with the ${\mathbf x}$-axis along $[100]$ and the
    ${\mathbf y}$-axis along $[001]$.  All three calculations were carried out
    for the experimental equilibrium structure and only the density from the
    valence states was considered.  The density was calculated at
    101$\times$101 grid points in (a), 151$\times$201 grid points in (b), and
    101$\times$151 grid points in (c).}
  \label{fig:pt2si_density}
\end{figure}

Each Si atom in $\alpha$-Pt$_2$Si has eight Pt nearest-neighbors at a distance
of 2.47~{\AA} [see Fig.~\ref{fig:crystal_structures}(a)].  In addition to four
Si nearest-neighbors, each Pt atom also has four Pt second-nearest-neighbors
at a distance of 2.79~{\AA} and two Pt third-nearest-neighbors at 2.98~\AA.
The Pt second-nearest-neighbors form two-dimensional (001) planes while the Pt
third-nearest-neighbors form linear $[001]$ chains.  The Pt
second-nearest-neighbor distance is very close to the nearest-neighbor
distance in pure fcc Pt and thus we might expect these two-dimensional planes
to exhibit evidence of metallic bonding.  This is in fact what we see, as
shown in Fig.~\ref{fig:pt2si_density}(a) which bears a strong resemblance to
the analogous plot in Fig.~\ref{fig:pt+si_density}(b).  However, this
approximately uniform increase in the charge density in the regions outside
the atomic cores is confined to the two-dimensional second-nearest-neighbor Pt
(001) planes.  In particular, there is little evidence of bonding (i.e. little
or no increase in the charge density relative to free atoms) along the
third-nearest-neighbor $[001]$ Pt chains.

In addition to the two-dimensional ``metallic'' bonding, we find strong
evidence of covalent bonding between the Pt and Si nearest-neighbors,
illustrated in Fig.~\ref{fig:pt2si_density}(b).  Unlike the case of pure Si
where the increase in charge density occurred between pairs of atoms, here the
density increase is localized between three atoms, two Pt and a Si.  For this
reason we refer to these as three-center covalent bonds.  We might even be
tempted to call these four-center bonds because there is a smaller increase in
the density, in between the two Pt atoms, which connects two of the
three-center bonds.  However, we note that the ${\mathbf x}$-axis in
Fig.~\ref{fig:pt2si_density}(b) is along the $[1\bar{1}0]$ direction and that
each of the Pt--Pt pairs in between two of the three-center bonds are also
located in one of the (001) planes which exhibit evidence of metallic bonding
[Fig.~\ref{fig:pt2si_density}(a)].  It would thus appear that rather than
four-center covalent bonds, a more appropriate description of the bonding in
$\alpha$-Pt$_2$Si would be three-center bonds interconnected by
two-dimensional metallic sheets.

The ${\mathbf y}$-axis of Fig.~\ref{fig:pt2si_density}(b) is along the $[111]$
direction which highlights two of the central Si atom's three-center bonds.
However, from Fig.~\ref{fig:crystal_structures}(a) we see that there are four
of these crystallographic directions and therefore a total of 8 of these
three-center bonds for each Si atom.  As noted above, the pair of Pt atoms
participating in a given three-center bond are second-nearest-neighbors
themselves.  Fig.~\ref{fig:pt2si_density}(c) shows that there is another set
of three-center bonds involving one Si atom and a pair of Pt atoms which are
third-nearest-neighbors oriented along the $[001]$ chains.  The ${\mathbf
x}$-axis in Fig.~\ref{fig:pt2si_density}(c) is along $[100]$ and the ${\mathbf
y}$-axis is along $[001]$.  There is little or no indication of an increase in
charge density along the Pt--Pt $[001]$ chains.  In addition to the two
three-center bonds in Fig.~\ref{fig:pt2si_density}(c), there are two more of
these bonds located in the plane obtained by a 90$^{\text{o}}$ rotation about
the $[001]$ axis [see Fig.~\ref{fig:crystal_structures}(a)], for a total of
four of these three-center bonds for each Si atom.

Thus we see that each Si atom in $\alpha$-Pt$_2$Si participates in twelve
three-center bonds, eight with Pt--Pt second-nearest-neighbors and four with
Pt--Pt third-nearest-neighbors, and that these three-center covalent bonds are
interconnected by two-dimensional second-nearest-neighbor Pt--Pt metallic
sheets.  Given the large increase in the number of bonds in $\alpha$-Pt$_2$Si
relative to pure Si we expect that each individual bond will be weaker than
one of the covalent bonds in Si.  However, taken as a whole and in terms of
the material strength, the more distributed nature of the bonding in
$\alpha$-Pt$_2$Si may indicate something closer in character to the pure
metallic bonding in fcc Pt.  This interpretation is supported by the
calculated elastic constants in Fig.~\ref{fig:elas_const_trends}, where the
non-volume-conserving elastic constants for $\alpha$-Pt$_2$Si are much closer
to those of fcc Pt as opposed to pure Si.  We address this issue in more
detail in Sec.~\ref{sec:force_field}.

\subsection{PtSi}
\label{sec:ptsi_density}

In the orthorhombic PtSi structure each Si atom has six Pt neighbors, with one
Pt at 2.41~{\AA}, two at 2.43~{\AA}, one at 2.52~{\AA}, and two at 2.64~{\AA}.
In view of the fact that the nearest-neighbor Pt--Si distance is 2.47~{\AA} in
$\alpha$-Pt$_2$Si it is perhaps not surprising that we find the two Pt
neighbors at 2.64~{\AA} appear to contribute little to the bonding in PtSi.
Each Si also has two Si fifth-nearest-neighbors at 2.84~{\AA} but again we
find little evidence of bonding between these atoms which is consistent with
the fact that the nearest-neighbor distance in pure Si is only 2.35~{\AA}.  In
addition to six Si neighbors at the same distances listed above, each Pt atom
also has two Pt neighbors at a sixth-nearest-neighbor distance of 2.87~{\AA}
and two more at a seventh-nearest-neighbor distance of 2.90~{\AA}.  These
distances are somewhat larger than the 2.77~{\AA} nearest-neighbor distance in
pure fcc Pt.

\begin{figure}[tbp]
  \centerline{\includegraphics[width=\linewidth]{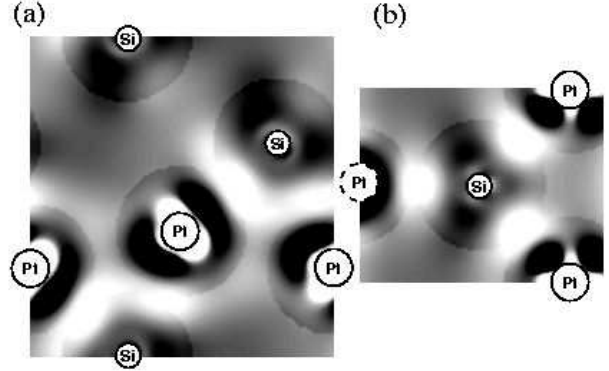}}
  \caption{Superposition of free atom densities subtracted from the fully
    self-consistent crystal density for orthorhombic PtSi.  The same 51
    contour levels and gray-scale are used as in
    Fig.~\protect\ref{fig:pt+si_density}.  In (a) we show the three-center
    Pt--Si--Pt covalent bonds with the ${\mathbf x}$-axis along $[\bar{1}00]$
    and the ${\mathbf y}$-axis along $[001]$.  The covalent Pt--Si bonds which
    connect atoms in adjacent ${\mathbf b}$-axis planes are shown in (b) with
    the ${\mathbf x}$-axis approximately along $[30\bar{4}$] and the ${\mathbf
    y}$-axis along $[010]$.  Both calculations were carried out for the
    experimental equilibrium structure and only the density from the valence
    states was considered.  The density was calculated at 101$\times$101 grid
    points in (a) and 201$\times$65 grid points in (b).  The leftmost Pt atom
    in (b) with the label contained inside a dotted circle is not actually
    located in the plane of the plot but is close enough that its influence
    can still be seen.}
  \label{fig:ptsi_density}
\end{figure}

The striking appearance of three-center bonds in $\alpha$-Pt$_2$Si is repeated
in orthorhombic PtSi, as shown in Fig.~\ref{fig:ptsi_density}(a).  As we see
in Fig.~\ref{fig:crystal_structures}(b), a convenient way to think of the PtSi
structure is as two alternating planes of atoms stacked along the ${\mathbf
b}$-axis.  Fig.~\ref{fig:ptsi_density}(a) shows the charge density difference
in one of these planes.  As in the case of $\alpha$-Pt$_2$Si
[Figs.~\ref{fig:pt2si_density}(b) and \ref{fig:pt2si_density}(c)] we see a
pileup of charge relative to the free atom density which is not localized
between a single pair of atoms but rather between one Si and two Pt atoms.
These Pt neighbors participating in the three-center bond are the first- and
third-nearest-neighbors of the Si atom and are at distances of 2.41~{\AA} and
2.52~{\AA}.  The two Pt atoms are themselves sixth-nearest-neighbors, with a
bond length of 2.87~{\AA}.  There appears to be a small increase in the charge
density between these two Pt atoms.  We note that the two different
three-center bonds shown in Fig.~\ref{fig:ptsi_density}(a) are equivalent by
symmetry.

The two second-neighbor Pt atoms of a given Si atom are located in adjacent
${\mathbf b}$-axis planes from the Si.  The charge density difference for
these bonds is shown in Fig.~\ref{fig:ptsi_density}(b), which indicates that
they are of the standard two-center variety.  In addition to these two-center
bonds, the plot also shows part of the bond with the first-neighbor Pt atom on
the left side of the figure.  In fact, the two second-neighbors as well as the
first- and third-neighbors form a very distorted tetrahedron around the
central Si atom.  The Pt--Si--Pt bond angles involving one Pt second-neighbor
and one third-neighbor are very nearly equal to the perfect tetrahedral angle
of 109.47$^{\text{o}}$ in pure Si, but the remaining four bond angles vary
considerably, ranging from 71$^{\text{o}}$ to 131$^{\text{o}}$.

There is very little evidence of an appreciable increase in the charge density
between the Pt and Si fourth-nearest-neighbors and the Si--Si fifth-neighbors,
as we mentioned above.  The Pt--Pt sixth-nearest-neighbors in
Fig.~\ref{fig:ptsi_density}(a) show some evidence of charge accumulation but
the Pt--Pt seventh-neighbors do not.  We thus see that there appears to be
only two sets of strong covalent bonds in orthorhombic PtSi, the three-center
Pt--Si--Pt bonds within a given ${\mathbf b}$-axis plane and the two-center
Pt--Si bonds between atoms in adjacent ${\mathbf b}$-axis planes, resulting in
a total of only three bonds per Si atom.  In this sense the bonding in PtSi
appears to be qualitatively much more similar to that in pure Si as compared
to pure Pt or even $\alpha$-Pt$_2$Si.  In particular, we are unable to
identify any concrete evidence in PtSi of a uniform increase in interstitial
charge density that might be associated with an element of metallic bonding.
We revisit this subject in Sec.~\ref{sec:ptsi_vff}.

\section{Valence force field models}
\label{sec:force_field}

In order to provide a more quantitative analysis of the trends in the elastic
constants as well as the various elements of the chemical bonding, we
construct simple valence force field models\cite{Har80,Tan99} to describe the
interatomic interactions for pure Pt, pure Si, and the two silicides.  In
these models the change in energy upon distorting the crystal, $\Delta E$, is
given as follows
\begin{equation}
  \label{eq:force_field}
  \begin{split}
    \Delta E = {\frac{N}{2}} n C_0 \left(\frac{\Delta V}{V}\right)^2
    &+ \frac{N}{2} \sum_i C_i \left(\frac{\Delta d_i}{d_i}\right)^2 \\
    &+ \frac{N}{2} \sum_{ij} K_{ij} \left(\Delta \theta_{ij}\right)^2,
  \end{split}
\end{equation}
where $N$ is the number of primitive cells in the crystal, $n$ is the number
of atoms in the primitive cell, $V$ is the volume, $\Delta d_i$ is the change
in the $i$th bond length , and $\Delta \theta_{ij}$ is the change in the bond
angle between the $i$th and $j$th bonds.  We determine the $C$ and $K$ force
constants by equating this expression for $\Delta E$ to corresponding elastic
constant expressions derived from Eq.~(\ref{eq:voigt_expansion}), examples of
which are given in Ref.~\onlinecite{Bec01}.  These coefficients are referred
to as force constants because Eq.~(\ref{eq:force_field}) could also be used to
anaylze the phonon spectrum and in this case, within a constant factor, the
coefficients play the role of Hooke's law force constants.  The factor of $n$
in the first term of Eq.~(\ref{eq:force_field}) is explicitly included so that
the resulting force constant $C_0$ represents the volume contribution per
atom, thus facilitating the comparison between materials with different
numbers of atoms in the primitive cell.  Similarly, the indices $i$ and $j$
are summed over all of the relevant bonds for each of the atoms in the
primitive cell (avoiding any double counting), which results in force
constants that represent the interaction strength for a single bond ($C_i$) or
bond angle ($K_{ij}$).  The volume term in Eq.~(\ref{eq:force_field}) is
needed for metals such as Pt and is reminiscent of the embedded-atom
method\cite{Daw83,Daw84} which has been successful in treating fcc metals.
Similarly, the angular terms are needed for covalently bonded systems such as
Si; such terms are a part of the Tersoff potential formulation\cite{Ter89}
which has been used successfully in semiconductor systems.  Both the volume
and the angular terms lead to deviations from the Cauchy relations\cite{Bor54}
which are strict equalities between various elastic constants that apply when
the interatomic interactions are purely pairwise [i.e. including only the
second term in Eq.~(\ref{eq:force_field})].

\subsection{Pt}
\label{sec:pt_vff}

In the case of fcc Pt we construct a two-parameter model, considering only the
nearest-neighbor bond length and a volume term, but no angular terms.  The
radial force constant $C_1$ can be obtained from the volume-conserving strains
corresponding to either $(c_{11}-c_{12})$ or $c_{44}$,
\begin{equation}
  \label{eq:pt_vff_c11-c12}
  {\frac{1}{2}} (c_{11}-c_{12}) = \frac{1}{v}\ {\frac{1}{4}} C_1,
\end{equation}
and
\begin{equation}
  \label{eq:pt_vff_c44}
  c_{44} = \frac{1}{v}\ {\frac{1}{2}} C_1,
\end{equation}
where $v = {\frac{1}{4}} a^3$ is the volume per atom.  Taken together, these
two equations provide an explanation for the fact that $(c_{11}-c_{12})$ and
$c_{44}$ for pure Pt are similar in magnitude in
Table~\ref{tab:pt+si_elastic}.  They also satisfy the cubic stability
requirements [Eq.~(\ref{eq:cubic_stability})] that $(c_{11}-c_{12}) > 0$ and
$c_{44} > 0$.

In order to facilitate comparison with the silicides where there is no
experimental data for the elastic constants, we also use the theoretically
determined elastic constants for fcc Pt.  For the purpose of internal
consistency we use the theoretical equilibrium volume as well.  The resulting
two values of $C_1$ obtained from Eqs.~(\ref{eq:pt_vff_c11-c12}) and
(\ref{eq:pt_vff_c44}) are 15.78~eV and 16.42~eV, respectively.  The fact that
the two numbers differ is an indication of the incompleteness of the
two-parameter model.  Use of the experimental elastic constants yields a
bigger difference but we nonetheless will use the average of these two values
for the purpose of comparing to the silicides,
\begin{equation}
  \label{eq:pt_vff_c1_avg}
  \bar{C_1} = v \left[(c_{11}-c_{12}) + c_{44}\right].
\end{equation}
Evaluated using the theoretical elastic constants we obtain $\bar{C_1}$ =
16.10~eV, while the experimental elastic constants correspond to a value of
16.95~eV.  We note that we could have eliminated the need to use the averaged
expression in Eq.~(\ref{eq:pt_vff_c1_avg}) by including additional force
constants but we prefer to maintain the conceptual simplicity of the
two-parameter model.  For example, including an angular interaction in
Eqs.~(\ref{eq:pt_vff_c11-c12}) and (\ref{eq:pt_vff_c44}) results in a small
and slightly negative angular force constant $K$ which is conceptually
unsatisfying.

The uniform expansion and compression represented by the bulk modulus $B_0$
can be used to obtain the following expression involving $C_0$ and $C_1$,
\begin{equation}
  \label{eq:pt_vff_b0}
  B_0 = {\frac{1}{3}} (c_{11} + 2 c_{12})
      = \frac{1}{v} \left(C_0 + {\frac{2}{3}} C_1 \right).
\end{equation}
Eqs.~(\ref{eq:pt_vff_c1_avg}) and (\ref{eq:pt_vff_b0}) together yield a value
of $C_0$ = 16.54~eV using the theoretical elastic constants, and 15.73~eV
using the experimental values.  The values of $C_0$ and $C_1$ obtained from
the theoretical elastic constants are listed in
Table~\ref{tab:force_constants}.  We note that the volume force constant has
approximately the same magnitude as the radial force constant and that both
are important in contributing to the large bulk modulus.  For completeness we
also give the expressions for $c_{11}$ and $c_{12}$ in terms of $C_0$ and
$C_1$,
\begin{equation}
  \label{eq:pt_vff_c11}
  c_{11} = \frac{1}{v} (C_0 + C_1),
\end{equation}
\begin{equation}
  \label{eq:pt_vff_c12}
  c_{12} = \frac{1}{v} \left(C_0 + {\frac{1}{2}} C_1\right).
\end{equation}
Eqs.~(\ref{eq:pt_vff_b0}--\ref{eq:pt_vff_c12}) explicitly satisfy the
stability requirement that $c_{12} < B_0 < c_{11}$ [Eq.~(\ref{eq:cubic_b0})].

The Cauchy relation for cubic crystals is that $c_{12} = c_{44}$.\cite{Bor54}
\ Using Eqs.~(\ref{eq:pt_vff_c44}) and (\ref{eq:pt_vff_c12}) we obtain the
following expression for the deviation from the cubic Cauchy relation,
\begin{equation}
  \label{eq:pt_vff_cauchy}
  (c_{12}-c_{44}) = \frac{1}{v} C_0.
\end{equation}
Thus we see that the large and positive deviation from the Cauchy relation in
pure Pt is due to a large volume contribution to $c_{12}$.  Moreover, the
presence of the volume contribution $C_0$ is responsible for the fact that
$c_{11}$, $c_{12}$, and $B_0$ are all significantly larger than the
volume-conserving elastic constants ${\frac{1}{2}}(c_{11} - c_{12})$ and
$c_{44}$.

\subsection{Si}
\label{sec:si_vff}

In the case of Si we also construct a two-parameter model but instead consider
only the nearest-neighbor bond length and the tetrahedral bond angle and set
all of the other $C$ and $K$ force constants to zero.  Since there are three
elastic constants and we allow only two force constants, we can check the
accuracy of the model.  The volume-conserving strain corresponding to
($c_{11}-c_{12}$) leaves the nearest-neighbor bond lengths unchanged to first
order in the distortion and thus only the angular force constant enters,
\begin{equation}
  \label{eq:si_vff_c11-c12}
  {\frac{1}{2}}(c_{11}-c_{12}) = \frac{1}{v}\ 2 K_{11},
\end{equation}
where $v = {\frac{1}{8}} a^3$ is the volume per atom.  Comparing this result
to Eq.~(\ref{eq:pt_vff_c11-c12}) we see that ${\frac{1}{2}}(c_{11} - c_{12})$
has a very different origin in Si as compared to fcc Pt, despite the fact that
the two values are approximately the same in Table~\ref{tab:pt+si_elastic}.

The volume-conserving $c_{44}$ strain in Si involves both radial and angular
distortions and can thus be used in conjunction with
Eq.~(\ref{eq:si_vff_c11-c12}) to determine $C_1$,
\begin{equation}
  \label{eq:si_vff_c44}
  c_{44}^{\text{frozen}}
    = \frac{1}{v} \left( {\frac{2}{9}} C_1 + {\frac{4}{9}} K_{11} \right),
\end{equation}
where $c_{44}^{\text{frozen}}$ corresponds to a pure $c_{44}$ strain, without
allowing for any internal relaxation.  This choice is convenient but not
essential and we can test how well the two force constants describe the final
remaining elastic constant.  Using the theoretically determined elastic
constants and equilibrium volume, Eqs.~(\ref{eq:si_vff_c11-c12}) and
(\ref{eq:si_vff_c44}) yield $C_1$ = 54.06~eV and $K_{11}$ = 3.13~eV, which are
listed in Table~\ref{tab:force_constants}.  Based on this analysis, the fact
that $(c_{11}-c_{12})$ and $c_{44}$ for Si are close in magnitude in
Table~\ref{tab:pt+si_elastic} is merely a coincidence having to do with the
specific values of the $C_1$ and $K_{11}$ force constants.

The final independent elastic constant is the bulk modulus $B_0$ which
corresponds to an isotropic expansion or compression and therefore only
involves radial but not angular distortions.  In addition, this distortion is
not volume-conserving and thus we could also have included the $C_0$ volume
term from Eq.~(\ref{eq:force_field}), which would not affect either of the
volume-conserving strains corresponding to Eqs.~(\ref{eq:si_vff_c11-c12}) and
(\ref{eq:si_vff_c44}), but would yield the following equation for the $B_0$
distortion,
\begin{equation}
  \label{eq:si_vff_b0}
  B_0 = {\frac{1}{3}} (c_{11} + 2 c_{12})
      = \frac{1}{v} \left( C_0 + {\frac{2}{9}} C_1 \right).
\end{equation}
Our two-parameter model has $C_0 \equiv 0$ and thus the extent to which $C_0$
obtained from Eq.~(\ref{eq:si_vff_b0}) deviates from zero provides a direct
measure of how well the two-parameter model is able to describe the elastic
constants.  Using the theoretical values determined here,
Eq.~(\ref{eq:si_vff_b0}) yields $C_0$ = $-$0.18~eV, which demonstrates that
the two-parameter model is indeed sufficiently accurate for describing the
elastic constants in Si.  For the sake of completeness we give the expressions
for $c_{11}$, $c_{12}$ and the deviation from the Cauchy relation
$(c_{12}-c_{44})$, including a volume contribution,
\begin{equation}
  \label{eq:si_vff_c11}
  c_{11} = \frac{1}{v} \left( C_0 + {\frac{2}{9}} C_1 + {\frac{8}{3}} K_{11} \right),
\end{equation}
\begin{equation}
  \label{eq:si_vff_c12}
  c_{12} = \frac{1}{v} \left( C_0 + {\frac{2}{9}} C_1 - {\frac{4}{3}} K_{11} \right),
\end{equation}
\begin{equation}
  \label{eq:si_vff_cauchy}
  (c_{12}-c_{44})^{\text{frozen}}
    = \frac{1}{v} \left( C_0 - {\frac{16}{9}} K_{11} \right).
\end{equation}

We can compare our two-parameter model to the one derived by
Harrison.\cite{Har80} \ His angular term has the identical form as ours and
his value of $K_{11} = 3.2$~eV differs from ours of 3.13~eV only because we
have used the theoretical elastic constants and equilibrium lattice constant
while he uses the experimental values.\cite{rem:notation} \ We derived the
value of the radial force constant $C_1$ using the frozen $c_{44}$ elastic
constant whereas Harrison derives his radial force constant from $B_0$.  The
two values would be identical if the value of $C_0$ derived from
Eq.~(\ref{eq:si_vff_b0}) were exactly zero.  The small deviation from zero, in
addition to the difference in the lattice constants used, leads to a small
difference between Harrison's value of $C_1 = 55.0$~eV and our value of
54.06~eV.

We have found that for Si the angular force constant $K_{11}$ is more than an
order of magnitude smaller than the radial force constant $C_1$
(Table~\ref{tab:force_constants}).  The angular interaction is nonetheless of
particular importance for two reasons.  The first is that the crystal would be
unstable in the absence of angular interactions since $(c_{11}-c_{12}) \equiv
0$ [Eq.~(\ref{eq:si_vff_c11-c12})] and $B_0 = c_{11} = c_{12}$
[Eqs.~(\ref{eq:si_vff_b0}--\ref{eq:si_vff_c12})] for purely radial
interactions, both of which violate the stability conditions discussed in
Sec.~\ref{sec:elastic_constants}.  The second reason is that the angular
interaction is responsible for the fact that the elastic constants do not obey
the Cauchy relation for a cubic crystal, $c_{12} = c_{44}$
[Eq.~(\ref{eq:si_vff_cauchy})].  Including a volume term but not an angular
interaction would still result in an unstable crystal since the
$(c_{11}-c_{12})$ distortion is volume-conserving and the dependence on $C_0$
is identical for $B_0$, $c_{11}$, and $c_{12}$.  In addition, a positive
volume contribution in the absence of an angular interaction could not account
for the fact that the deviation from the Cauchy relation is negative $(c_{12}
< c_{44})$.  Thus we see that Eq.~(\ref{eq:si_vff_cauchy}) and the fact that
$K_{11}$ is small combine to provide an explanation for the fact that $c_{12}$
is smaller than $c_{44}$, but only by a relatively small amount.  This is in
sharp contrast to pure Pt where the deviation from the Cauchy relation
[Eq.~(\ref{eq:pt_vff_cauchy})] arises from $C_0$ and is large and positive,
resulting in a value of $c_{12}$ which is more than four times larger than in
Si.  More generally, the absence of a volume contribution in Si is responsible
for the fact that $c_{12}$ and $B_0$ are similar in magnitude to the
volume-conserving elastic constants ${\frac{1}{2}}(c_{11} - c_{12})$ and
$c_{44}$, in contrast to the case of Pt.

From the force constants listed in Table~\ref{tab:force_constants} we see that
$C_1$ for Si is more than a factor of three times larger than for fcc Pt which
is consistent with the presence of strong covalent bonds in Si and distributed
metallic bonding in Pt.  In addition, the volume contribution $C_0$ is equally
important in terms of the metallic bonding in Pt but plays no role in Si.  The
influence on the elastic constants of these qualitative differences in the
chemical bonding are clearly illustrated by comparing
Eqs.~(\ref{eq:pt_vff_b0}) and (\ref{eq:si_vff_b0}) for the bulk moduli in Pt
and Si, respectively.  We see that the geometry coefficient of $C_1$ is three
times larger for Pt than for Si, reflecting the difference in the
nearest-neighbor coordination and nearly compensating for the difference in
the magnitudes of the two force constants.  Given that $C_0$ and $C_1$ are
approximately the same in Pt, we see from Eq.~(\ref{eq:pt_vff_b0}) that the
volume contribution to $B_0$ is approximately 50\% larger than the
contribution from $C_1$.  In the case of Si the $C_0$ force constant is
essentially zero and this difference accounts for most of the difference in
the magnitudes of $B_0$ between Pt and Si.  The pre-factors of $1/v$ account
for the remaining difference since the volume per atom $v$ is 30\% larger in
Si.

We therefore see that the presence or absence of metallic bonding, as
reflected in the $C_0$ and $C_1$ force constants, is intimately connected to
the magnitudes of $B_0$.  Similar analyses can be used to explain the fact
that $c_{11}$ and $c_{12}$ are also larger in Pt, the predominant reason being
the presence of a large volume contribution (or equivalently, metallic
bonding) in Pt but not Si.  Conversely, the elastic constants corresponding to
both of the volume-conserving distortions in Pt, ${\frac{1}{2}} (c_{11} -
c_{12})$ and $c_{44}$, are approximately the same as in Si, indicating that
they are less sensitive to the differences in chemical bonding for these two
materials.  These differing trends in the volume-conserving versus
non-volume-conserving elastic constants were already noted in
Sec.~\ref{sec:elastic_constants} and are illustrated in
Fig.~\ref{fig:elas_const_trends}.  The volume force constant $C_0$ is included
in the figure on the same scale as the elastic constants by dividing by the
appropriate volume per atom $v$ (note that $\frac{1}{v}C_0$ is precisely the
combination that enters all of the expressions for the non-volume-conserving
elastic constants).

\subsection{{\boldmath$\alpha$}-Pt{\boldmath$_2$}Si}
\label{sec:pt2si_vff}

In order to provide a more quantitative description of the chemical bonding in
$\alpha$-Pt$_2$Si we describe the interatomic interactions using a valence
force field model, just as we did for pure fcc Pt and pure Si.  In view of the
analysis of the valence charge density in Sec.~\ref{sec:pt2si_density}, we
include first-, second-, and third-nearest-neighbor radial force constants as
well as a volume term.  We also consider some of the angular interactions.  In
keeping with our neglect of angular interactions in pure Pt we also neglect
the bond angles between any two Pt--Pt bonds, both in the two-dimensional
(001) metallic sheets and the $[001]$ Pt chains.  In consideration of the
three-center bonds discussed earlier, we include both the Pt--Si--Pt and the
Si--Pt--Pt bond angles relevant to the three-center bonds involving one Si
atom and two second-neighbor Pt atoms.  However, we neglect the bond angles
relevant to the three-center bonds involving one Si and two third-neighbor Pt
atoms.  This choice is based on the expectation that the strength of the
angular interactions will generally be smaller than that of the radial
interactions and that the three-center bond involving two Pt second-neighbors
is stronger than the one involving two third-neighbors.  We thus have six
force constants which can be fit to the six elastic constants.  For the sake
of convenience we fit the force constant expressions to the ``frozen'' elastic
constants, where no internal relaxations were carried out.  This choice is not
essential and need not be considered an additional approximation because the
resulting force constants could be used to directly calculate the internal
relaxations.

The volume-conserving strains corresponding to $(c_{11}-c_{12})$ and $c_{44}$
both depend only on the first-nearest-neighbor Pt--Si radial force constant
$C_1$ since the second- and third-nearest-neighbor Pt--Pt bond lengths are
left unchanged to first order.  In addition, $(c_{11}-c_{12})$ depends on the
Si--Pt--Pt bond angle but not the Pt--Si--Pt bond angle, while $c_{44}$
depends on both.  We label the force constant for the Pt--Si--Pt bond angle as
$K_{11}$ because it is the angle between two first-neighbor bonds.  Similarly
we label the Si--Pt--Pt force constant as $K_{12}$.  Equating the elastic
constant [Eq.~(\ref{eq:voigt_expansion})] and force constant
[Eq.~(\ref{eq:force_field})] expressions for the change in the energy and
using the theoretical lattice constants from Table~\ref{tab:atomic_structure},
we obtain the following two equations,
\begin{equation}
  \label{eq:pt2si_vff_c11-c12}
  {\frac{1}{2}}(c_{11}-c_{12})^{\text{frozen}}
    = \frac{1}{v} (0.2685\ C_1 + 1.1553\ K_{12}),
\end{equation}
and
\begin{equation}
  \label{eq:pt2si_vff_c44}
  c_{44}^{\text{frozen}}
    = \frac{1}{v} (0.3092\ C_1 + 0.2874\ K_{11} + 0.8089\ K_{12}),
\end{equation}
where $v = 0.2529\ a^3$ is the volume per atom.  We note that in these
equations and all of those that follow, the numerical coefficients are simply
geometrical factors containing various combinations of the $a$ and $c$ lattice
constants.  Examining the geometry coefficients of $C_1$ in
Eqs.~(\ref{eq:pt2si_vff_c11-c12}) and (\ref{eq:pt2si_vff_c44}), together with
the expectation that the angular force constants will be significantly smaller
in magnitude than $C_1$, we see that these two equations provide a natural
explanation for why $\frac{1}{2} (c_{11}-c_{12})$ is similar in magnitude but
slightly smaller than $c_{44}$ in Table~\ref{tab:silicides_elastic}.  We also
see that two of the stability requirements [Eq.~(\ref{eq:pt2si_stability})],
$(c_{11}-c_{12}) > 0$ and $c_{44} > 0$, are explicitly satisfied.

The volume-conserving strain corresponding to $c_{66}$ changes the
second-neighbor Pt--Pt bond length, leaving the other two bond lengths
unchanged to first order.  This strain also modifies the two bond angles
yielding,
\begin{equation}
  \label{eq:pt2si_vff_c66}
  c_{66} = \frac{1}{v} \left( {\frac{1}{3}} C_2 + 1.2394\ K_{11}
    + 0.6197\ K_{12} \right).
\end{equation}
Thus we can see from Eqs.~(\ref{eq:pt2si_vff_c44}) and
(\ref{eq:pt2si_vff_c66}) that since $c_{66}$ in
Table~\ref{tab:silicides_elastic} is somewhat more than two times larger than
$c_{44}$, we expect that the second-neighbor Pt--Pt force constant $C_2$ must
be approximately two times larger than the first-neighbor Pt--Si force
constant $C_1$.  We will in fact find this to be the case.
Eq.~(\ref{eq:pt2si_vff_c66}) also satisfies the stability requirement that
$c_{66} > 0$.

The final remaining volume-conserving strain corresponding to
$(c_{11}+c_{33}-2c_{13})$ changes all of the first-, second-, and
third-neighbor bond lengths, as well as the two bond angles,
\begin{multline}
  \label{eq:pt2si_vff_c11+c33-2c13}
  {\frac{1}{4}}(c_{11}+c_{33}-2c_{13})^{\text{frozen}} \\
  = \frac{1}{v} \bigg(0.0687\ C_1 + {\frac{1}{12}} C_2 + {\frac{1}{6}} C_3 \\
    + 0.3723\ K_{11} + 0.4750\ K_{12}\bigg) .
\end{multline}
We note that ${\frac{1}{4}}(c_{11}+c_{33}-2c_{13}) > 0$ as required for
mechanical stability, and that it is similar in magnitude to $\frac{1}{2}
(c_{11}-c_{12})$ and $c_{44}$ in Table~\ref{tab:silicides_elastic}.

The uniform expansion and compression corresponding to the bulk modulus $B_0$
changes the volume and all of the bond lengths but leaves the bond angles
fixed,
\begin{equation}
  \label{eq:pt2si_vff_b0}
  \begin{split}
    B_0 &= {\frac{1}{9}}(2c_{11} + c_{33} + 2c_{12} + 4c_{13}) \\
        &= \frac{1}{v} \left(
      C_0 + {\frac{8}{27}} C_1 + {\frac{4}{27}} C_2 + {\frac{2}{27}} C_3\right).
\end{split}
\end{equation}
The final two equations resulting from the $c_{11}$ and $c_{33}$ strains both
include a contribution from a change in the volume,
\begin{equation}
  \label{eq:pt2si_vff_c11}
  \begin{split}
    c_{11}^{\text{frozen}} = \frac{1}{v} \bigg(& C_0 + 0.5370\ C_1 \\
      &+ {\frac{1}{3}} C_2 + 0.1655\ K_{11} + 1.2381\ K_{12} \bigg),
  \end{split}
\end{equation}
and
\begin{equation}
  \label{eq:pt2si_vff_c33}
  \begin{split}
    c_{33} = \frac{1}{v} \bigg(& C_0 + 0.3560\ C_1 + {\frac{2}{3}} C_3 \\
    &+ 0.6619\ K_{11} + 0.3310\ K_{12} \bigg).
  \end{split}
\end{equation}
Similarly, the equations for $c_{12}$ and $c_{13}$ are
\begin{equation}
  \label{eq:pt2si_vff_c12}
  c_{12}^{\text{frozen}} = \frac{1}{v} \left( C_0 + {\frac{1}{3}} C_2
                      + 0.1655\ K_{11} - 1.0726\ K_{12} \right),
\end{equation}
and
\begin{equation}
  \label{eq:pt2si_vff_c13}
  c_{13} = \frac{1}{v} (C_0 + 0.3092\ C_1 - 0.3310\ K_{11} - 0.1655\ K_{12}).
\end{equation}
Eqs.~(\ref{eq:pt2si_vff_b0})--(\ref{eq:pt2si_vff_c13}) explicitly satisfy the
stability requirements [Eq.~(\ref{eq:pt2si_b0_stability})] that $B_0 <
{\frac{1}{3}}(2 c_{11} + c_{33})$ and $B_0 > {\frac{1}{3}}(c_{12} + 2
c_{13})$.  We note that in contrast to the case of Si, all of the stability
requirements in Eq.~(\ref{eq:pt2si_stability}) would be satisfied even for
purely radial interactions (i.e. no angular interactions).

Eqs.~(\ref{eq:pt2si_vff_c11-c12})--(\ref{eq:pt2si_vff_c11+c33-2c13}),
(\ref{eq:pt2si_vff_c11}), and (\ref{eq:pt2si_vff_c33}) represent six linearly
independent equations in the six unknown force constants.  Solving this linear
system of equations yields the force constants listed in
Table~\ref{tab:force_constants} for $\alpha$-Pt$_2$Si.  The volume force
constant $C_0$ is only 16\% smaller than in pure Pt.  This is consistent with
the presence of two-dimensional metallic sheets in $\alpha$-Pt$_2$Si and the
fact that there are a large number of distributed three-center bonds all
interconnected by these sheets.  The first neighbor Pt--Si force constant
$C_1$ is nearly four times smaller than $C_1$ in pure Si.  This large
reduction results from the fact that each Si atom in $\alpha$-Pt$_2$Si has
eight Pt nearest-neighbors and participates in twelve different three-center
bonds.  Conversely, the second-neighbor Pt--Pt force constant $C_2$ is 60\%
larger than the corresponding $C_1$ force constant in pure fcc Pt, despite the
fact that the two Pt--Pt bond lengths are very nearly the same.  We can
understand this result because each Pt atom in pure Pt has twelve nearest
neighbors while each Pt in $\alpha$-Pt$_2$Si has only four Pt second-neighbors
and two Pt third-neighbors.  Moreover, the Pt atoms in the silicide
participate in covalent three-center bonds in addition to the metallic bonding
within the two-dimensional sheets.  The distributed nature of these bonds and
the large number of them in the primitive cell are both consistent with the
fact that $C_2$ is still a factor of two smaller than $C_1$ in pure Si.  We
found little evidence of an increase in the electronic charge density between
the Pt--Pt third-neighbors and this is reflected in the fact that $C_3$ is
more than three times smaller than $C_2$.  We also find that the angular force
constants $K_{11}$ and $K_{12}$ are similar in magnitude to the $K_{11}$ force
constant in pure Si.  These angular terms play an important but less crucial
role in the silicide as compared to pure Si.

Having determined the values of the individual force constants we can now use
them to understand the trends in the elastic constants.  For example, the two
Cauchy relations for tetragonal crystals are that $c_{12} = c_{66}$ and
$c_{13} = c_{44}$.\cite{Bor54} \ Using Eqs.~(\ref{eq:pt2si_vff_c66}) and
(\ref{eq:pt2si_vff_c12}) the deviation from the first Cauchy relation is given
by
\begin{equation}
  \label{eq:pt2si_vff_cauchy_c66}
  (c_{12}-c_{66})^{\text{frozen}}
    = \frac{1}{v} (C_0 - 1.0739\ K_{11} - 1.6923\ K_{12}).
\end{equation}
Similarly, the deviation from the second Cauchy relation is
\begin{equation}
  \label{eq:pt2si_vff_cauchy_c44}
  (c_{13}-c_{44})^{\text{frozen}}
    = \frac{1}{v} (C_0 - 0.6184\ K_{11} - 0.9744\ K_{12}).
\end{equation}
As in the case of pure Pt [Eq.~(\ref{eq:pt_vff_cauchy})], it is the presence
of the volume interaction which produces a positive deviation from the Cauchy
relations.  The angular interactions provide a negative contribution to
Eqs.~(\ref{eq:pt2si_vff_cauchy_c66}) and (\ref{eq:pt2si_vff_cauchy_c44}), just
as they did for pure Si [Eq.~(\ref{eq:si_vff_cauchy})].  From the geometry
coefficients of the angular terms we see that $(c_{13}-c_{44})$ must be larger
in magnitude than $(c_{12}-c_{66})$.  The net result is that the deviations
from the Cauchy relations for $\alpha$-Pt$_2$Si are still positive but are
factors of 2--3 smaller than the deviation in pure Pt.

In pure Pt we found that the volume-conserving elastic constants were all
significantly smaller than the others and that this was due predominantly to
the presence of a large volume contribution $C_0$.  We find the same trend in
$\alpha$-Pt$_2$Si with ${\frac{1}{2}}(c_{11} - c_{12})$, $c_{44}$, and
$\frac{1}{4} (c_{11} + c_{33} - 2 c_{13})$ all being similar in magnitude and
smaller than all the remaining elastic constants (see
Fig.~\ref{fig:elas_const_trends}).  The notable exception to this trend is
$c_{66}$.  In conjunction with Eq.~(\ref{eq:pt2si_vff_c66}) we already noted
that the large value of $c_{66}$ in relation to the other volume-conserving
elastic constants is due primarily to the fact that the second-neighbor Pt--Pt
force constant $C_2$ is a factor of two larger than the first neighbor Pt--Si
force constant $C_1$.  This result is in turn directly related to the presence
of the network of three-center bonds interconnected by two-dimensional
metallic sheets.  We also saw that mechanical stability requires $c_{12} <
c_{11}$, $c_{13} < {\frac{1}{2}}(c_{11} + c_{33})$, $B_0 < {\frac{1}{3}}(2
c_{11} + c_{33})$, and $B_0 > {\frac{1}{3}}(c_{12} + 2 c_{13})$ and that the
deviations from the Cauchy relations are positive.  The remaining variations
among the six elastic constants in Table~\ref{tab:silicides_elastic} are
determined by the detailed dependence on the various force constants as
described above.

One interesting example is that $c_{13}$ is essentially identical to $c_{66}$.
Comparing Eqs.~(\ref{eq:pt2si_vff_c66}) and (\ref{eq:pt2si_vff_c13}) we see
that in the case of $c_{13}$ positive volume and first-neighbor radial terms
are partially counterbalanced by negative angular contributions whereas
$c_{66}$ corresponds to a volume-conserving distortion and has positive
angular contributions.  In addition, the $c_{13}$ distortion changes the
first-neighbor bond lengths, leaving the others fixed, while the $c_{66}$
distortion changes the second-neighbor bond lengths, leaving the others fixed.
We have already noted that $c_{66}$ is anomalously large in comparison to
$c_{44}$ predominantly because $C_2$ is twice as large as $C_1$.  The elastic
constant $c_{12}$ is larger than $c_{13}$ for the same reason, thus explaining
how it is at least possible for the volume-conserving elastic constant
$c_{66}$ to be similar in magnitude to the non-volume-conserving elastic
constant $c_{13}$.  In summary, while we are able to explain the overall
magnitudes of the individual elastic constants, we are forced to conclude that
the specific equality of $c_{13}$ and $c_{66}$ in
Table~\ref{tab:silicides_elastic} depends on the precise values of the
individual force constants and is therefore simply accidental.

We saw in Eq.~(\ref{eq:pt_vff_b0}) for the bulk modulus of Pt that the volume
contribution represented 60\% of the total, with the contribution from the
radial interaction making up the rest.  The same approximate 60:40 split
between the volume and radial contributions applies to the expression for the
bulk modulus of $\alpha$-Pt$_2$Si in Eq.~(\ref{eq:pt2si_vff_b0}).  In
addition, the volume per atom $v$ is nearly the same in the two materials.
Thus we see that the 16\% reduction in $C_0$ for $\alpha$-Pt$_2$Si relative to
Pt, combined with a similar reduction in the overall radial contribution,
leads to a bulk modulus which is approximately 20\% smaller in
$\alpha$-Pt$_2$Si.  As we noted previously, there is no volume contribution in
Si where the bulk modulus is a factor of 2--3 smaller.  Conversely, the
volume-conserving elastic constant $c_{44}$ is similar in magnitude for all
three materials.  Comparing Eqs.~(\ref{eq:pt_vff_c44}), (\ref{eq:si_vff_c44}),
and (\ref{eq:pt2si_vff_c44}) we see that in pure Pt $c_{44}$ arises solely
from $C_1$ whereas in pure Si and $\alpha$-Pt$_2$Si it arises from a
combination of $C_1$ and angular contributions.  In Si the split is 90:10,
radial to angular, while in the silicide the split is only 60:40 since $C_1$
is a factor of four smaller.  In addition, the volume per atom is 30\% larger
in Si than in the other two materials.  The remaining elastic constants for
$\alpha$-Pt$_2$Si can be similarly analyzed in relation to those of pure Pt
and pure Si.

\subsection{PtSi}
\label{sec:ptsi_vff}

Once again we construct a valence force field model to describe the chemical
bonding and elastic constant trends in PtSi.  In keeping with the discussion
of the valence charge density in Sec.~\ref{sec:ptsi_density}, we include
first-, second-, and third-neighbor Pt--Si radial force constants (labeled
$C_1$, $C_2$, and $C_3$) as well as sixth- and seventh-neighbor Pt--Pt radial
force constants (labeled $C_6$ and $C_7$).  We also include a volume term
($C_0$) and three Pt--Si--Pt angular force constants.  The angular force
constants are labeled $K_{13}$, corresponding to the bond angle between first-
and third-neighbor Pt--Si bonds, $K_{22}$, corresponding to the bond angle
between two second-neighbor Pt--Si bonds, and $K_{23}$, corresponding to the
bond angle between second- and third-neighbor Pt--Si bonds.  These three bond
angles are the ones we have found to be most important and are the ones which
correspond to the distorted tetrahedral Pt--Si--Pt angles described in
Sec.~\ref{sec:ptsi_density}.  The fourth and last of these angles is
represented by the force constant $K_{12}$ but we found it to be unimportant
and have not included it in the analysis presented here.  Part of the reason
for this finding may be that this bond angle is 131.72$^{\text{o}}$ which is
quite different from the perfect tetrahedral angle of 109.47$^{\text{o}}$.  We
thus have nine force constants which can be fit to the nine elastic constants.
As in the case of $\alpha$-Pt$_2$Si we fit the force constant expressions to
the ``frozen'' elastic constants out of convenience, but this choice is not
essential because the relaxations could be calculated from the resulting
model.

\begin{figure}[tbp]
  \centerline{\includegraphics[width=0.8\linewidth]{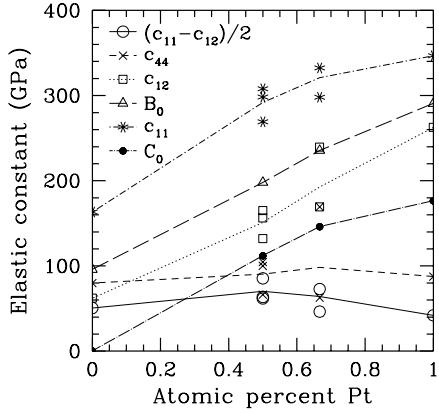}}
  \caption{Trends in the elastic constants as a function of atomic percent Pt
    for pure cubic-diamond-phase Si, orthorhombic PtSi, tetragonal
    $\alpha$-Pt$_2$Si, and fcc Pt.  The different curves correspond to the
    average values of different classes of the individual elastic constants,
    as specified in the legend.  For example, in the case of the dotted-line
    curve labeled as $c_{12}$, the line passes through ${\frac{1}{3}}(c_{12} +
    c_{13} + c_{23})$ in the case of PtSi and through ${\frac{1}{3}}(c_{12} +
    2 c_{13})$ for $\alpha$-Pt$_2$Si ($c_{13} = c_{23}$ for tetragonal
    crystals), while the open squares show the actual values of $c_{12}$,
    $c_{13}$ and $c_{23}$, as appropriate for each material.  The $C_0$ force
    constant curve is scaled by the inverse of the volume per atom in order to
    be able to plot it on the same scale as the elastic constants.  The
    significance of $C_0$ in connection to the elastic constants is discussed
    in Sec.~\protect\ref{sec:force_field}.}
  \label{fig:elas_const_trends}
\end{figure}

Most of the expressions for the elastic constants in terms of the force
constants involve all of the radial and angular terms and thus there is not
much to be learned by writing them down.  Two exceptions are the
volume-conserving strains corresponding to $c_{44}$ and $c_{66}$ which depend
only on the second-neighbor Pt--Si and seventh-neighbor Pt--Pt radial force
constants, as well as the angular force constant $K_{23}$.  Using the
theoretically determined structural parameters from
Table~\ref{tab:atomic_structure} we obtain the following two expressions,
\begin{equation}
  \label{eq:ptsi_vff_c44}
  c_{44}^{\text{frozen}}
    = \frac{1}{v} (0.1601\ C_2 + 0.1182\ C_7 + 0.3279\ K_{23}),
\end{equation}
and
\begin{equation}
  \label{eq:ptsi_vff_c66}
  c_{66}^{\text{frozen}}
    = \frac{1}{v} (0.0882\ C_2 + 0.00002\ C_7 + 0.1750\ K_{23}),
\end{equation}
where $v = 0.08464\ a^3$ is the volume per atom.  The force constant $C_7$
will turn out to be small and thus we can see from
Eqs.~(\ref{eq:ptsi_vff_c44}) and (\ref{eq:ptsi_vff_c66}) that
$c_{44}^{\text{frozen}}$ is approximately a factor of two larger than
$c_{66}^{\text{frozen}}$ purely because of geometrical factors.  In addition,
the mechanical stability requirements that $c_{44} > 0$ and $c_{66} > 0$ are
satisfied by a combination of radial and angular terms.  However, as in the
case of $\alpha$-Pt$_2$Si, the angular terms are not essential with regard to
stability since the crystal would still be stable under purely radial
interactions.  It turns out that this circumstance is true for all of the
stability requirements in Eq.~(\ref{eq:ptsi_stability}).  We also note that
all of the volume-conserving elastic constants,
${\frac{1}{4}}(c_{11}+c_{22}-2c_{12})$,
${\frac{1}{4}}(c_{11}+c_{33}-2c_{13})$,
${\frac{1}{4}}(c_{22}+c_{33}-2c_{23})$, $c_{44}$, $c_{55}$, and $c_{66}$ are
similar in magnitude and smaller than the other non-volume-conserving elastic
constants (see Fig.~\ref{fig:elas_const_trends}).  The primary exception is
$c_{44}^{\text{frozen}}$, although including the effect of relaxation brings
it in line with the other volume-conserving constants.

Solving the linear system of nine equations in the nine unknown force
constants, we obtain the values listed in Table~\ref{tab:force_constants}.
The volume force constant $C_0$ is nearly 40\% smaller than in pure Pt and
25\% smaller than in $\alpha$-Pt$_2$Si.  Nonetheless, the value is still
sizeable and perhaps somewhat surprising given that we found no evidence of
metallic-type bonding in our analysis of the charge density in
Sec.~\ref{sec:ptsi_density}.  The first- and second-neighbor Pt--Si radial
force constants are quite large and nearly as large as the first-neighbor
Si--Si force constant in pure Si.  This result is consistent with the fact
that we found only a small number of two- and three-center bonds for each Si
atom in PtSi.  This small number of bonds means that each bond is relatively
strong, as is the case in pure Si, but in contrast to the situation in
$\alpha$-Pt$_2$Si where the Pt--Si $C_1$ force constant is more than a factor
of three smaller.  The fact that $C_1$ and $C_2$ in PtSi are still smaller
than $C_1$ is Si may be due to the fact that the bond angles in PtSi are
considerable distorted away from the perfect tetrahedral angle.  The Pt--Si
$C_3$ force constant in PtSi is approximately a factor of four smaller than
$C_1$ and $C_2$ which may be due in part to the correspondingly longer bond
length.

The Pt--Pt sixth-neighbor force constant $C_6$ is larger than $C_1$ in pure Pt
which is likely due to the fact that this interaction contributes to the
three-center bonds in PtSi.  However, $C_6$ is 30\% smaller than the
corresponding Pt--Pt $C_2$ force constant in $\alpha$-Pt$_2$Si, reflecting the
longer bond length and the lack of two-dimensional metallic sheets in PtSi.
Although the seventh-neighbor Pt--Pt bond length in PtSi is only 0.03~{\AA}
larger than the sixth-neighbor bond length, the seventh-neighbor bond does not
participate in any three-center bonds and we found little evidence of any
increase in the charge density.  It is thus not surprising that $C_7$ is more
than a factor of four smaller than $C_6$.

We find that the angular interactions are sizeable in PtSi, as they were in
$\alpha$-Pt$_2$Si.  However, in PtSi these interactions show a wider variation
in magnitude, with $K_{23}$ being more than an order of magnitude larger than
$K_{13}$.  We can understand the variation in these Pt--Si--Pt force constants
by looking at the sizes of the bond angles themselves.  $K_{13}$ corresponds
to a bond angle of 71.09$^{\text{o}}$ which is very far from the perfect
tetrahedral angle of 109.47$^{\text{o}}$.  The bond angle associated with
$K_{22}$ is a lot closer, having a value of 94.64$^{\text{o}}$, resulting in a
larger force constant.  The largest angular force constant is $K_{23}$ with
the corresponding bond angle of 109.75$^{\text{o}}$ being nearly identical to
the perfect tetrahedral angle.  While the trend in the angular force constants
in PtSi is understandable in terms of the deviation relative to the pure
tetrahedral angle, the large magnitude of $K_{23}$ in comparison to the
angular force constants in pure Si and $\alpha$-Pt$_2$Si is unexpected.  The
angular interactions appear to be of greater importance in PtSi than they were
in $\alpha$-Pt$_2$Si.  An attempt to fit the elastic constants of PtSi using a
valence force field model including only radial interactions plus a volume
term resulted in nonsensical values for these force constants.  A sensible fit
was only achieved after including angular terms.

We can now examine some of the trends in the elastic constants of PtSi using
the calculated force constants.  In particular, the Cauchy relations for an
orthorhombic crystal are that $c_{12} = c_{66}$, $c_{13} = c_{55}$, and
$c_{23} = c_{44}$.\cite{Bor54} \ The expressions for the deviations from these
Cauchy relations are as follows,
\begin{equation}
  \label{eq:ptsi_vff_cauchy_c66}
  (c_{12}-c_{66})^{\text{frozen}}
    = \frac{1}{v} (C_0 - 0.1765\ K_{22} - 0.0860\ K_{23}),
\end{equation}
\begin{equation}
  \label{eq:ptsi_vff_cauchy_c55}
  (c_{13}-c_{55})^{\text{frozen}}
    = \frac{1}{v} (C_0 - 0.4475\ K_{13} - 0.3453\ K_{23}),
\end{equation}
\begin{equation}
  \label{eq:ptsi_vff_cauchy_c44}
  (c_{23}-c_{44})^{\text{frozen}}
    = \frac{1}{v} (C_0 - 0.3202\ K_{22} - 0.4544\ K_{23}).
\end{equation}
As in the case of $\alpha$-Pt$_2$Si the volume interaction makes a positive
contribution to the deviations from the Cauchy relations while the angular
interactions make a negative contribution.  The geometry coefficients for the
angular terms in
Eqs.~(\ref{eq:ptsi_vff_cauchy_c66})--(\ref{eq:ptsi_vff_cauchy_c44}) are
smaller than for $\alpha$-Pt$_2$Si in Eqs.~(\ref{eq:pt2si_vff_cauchy_c66}) and
(\ref{eq:pt2si_vff_cauchy_c44}), reflecting the smaller multiplicity of the
bond angles in PtSi.  This reduction is more than compensated by the larger
magnitude of the force constants in PtSi, particularly $K_{23}$.  The volume
per atom $v$ is similar in the two silicides but the magnitude of $C_0$ is
smaller in PtSi.  The combined effect of the smaller $C_0$ and the larger
$K_{23}$ is that the deviations from the Cauchy relations in
Eqs.~(\ref{eq:ptsi_vff_cauchy_c66})--(\ref{eq:ptsi_vff_cauchy_c44}) are still
positive but approximately 30\% smaller on average than in $\alpha$-Pt$_2$Si.
This conclusion remains true for the relaxed elastic constants, although the
specific numerical details are changed.  For example, the larger geometry
coefficients of $K_{22}$, and especially $K_{23}$, in
Eq.~(\ref{eq:ptsi_vff_cauchy_c44}) result in a very small deviation from the
third Cauchy relation $(c_{23}-c_{44})^{\text{frozen}}$ for the frozen elastic
constants.  When relaxation is included $c_{44}$ drops by 29\% while $c_{23}$
increases by 8\%, resulting in a significantly larger deviation.  However,
$(c_{13}-c_{55})$ becomes much smaller so that on average the deviations are
still approximately 30\% smaller in PtSi.

The requirements of mechanical stability in Eqs.~(\ref{eq:ptsi_stability}) and
(\ref{eq:ptsi_b0_stability}) constrain the elastic constants by requiring that
$c_{12} < {\frac{1}{2}}(c_{11} + c_{22})$, $c_{13} < {\frac{1}{2}}(c_{11} +
c_{33})$, $c_{23} < \frac{1}{2} (c_{22} + c_{33})$, $B_0 <
{\frac{1}{3}}(c_{11} + c_{22} + c_{33})$, and $B_0 > {\frac{1}{3}}(c_{12} +
c_{13} + c_{23})$.  However, there are additional trends among the elastic
constants.  We have already noted that the volume-conserving elastic constants
in PtSi are all smaller than those where the corresponding distortion does not
conserve volume.  The predominant reason for this occurrence is the presence
of the positive volume contribution $C_0$, just as it was in the case of pure
Pt and in $\alpha$-Pt$_2$Si (see Fig.~\ref{fig:elas_const_trends}).  The
positive deviations from the Cauchy relations in
Eqs.~(\ref{eq:ptsi_vff_cauchy_c66})--(\ref{eq:ptsi_vff_cauchy_c44}) provide
specific examples of this trend.  We noted above that the relatively large
value of $C_0$ in PtSi seemed surprising given the lack of evidence for
metallic bonding in the charge density.  In fact, it appeared that the
chemical bonding in PtSi was much more similar to that in pure Si than in
either $\alpha$-Pt$_2$Si or pure Pt.  However, the trends in the elastic
constants of PtSi, the positive deviations from the Cauchy relations and the
smaller values of the volume-conserving elastic constants, are much more
similar to those in the materials that do exhibit direct evidence of metallic
bonding, thus requiring a sizeable $C_0$ volume contribution in PtSi as well.
This conclusion is not one that we would have reached based on the charge
density alone, thus demonstrating the need for care when examining such
qualitative characteristics.  By contrast, the analysis of the elastic
constants using a valence force field model has allowed a more quantitative
description of the chemical bonding.  We note that the finding of both
metallic and covalent components to the bonding in PtSi as well as
$\alpha$-Pt$_2$Si indicates a strong similarity between these two materials
and may also be connected with the fact that the heats of formation for the
two are very nearly the same.\cite{Bec01}

Finally, we examine how the elastic constants of PtSi fit into the trends
between the different materials studied here.  The expression for the bulk
modulus in PtSi is
\begin{multline}
  \label{eq:ptsi_vff_b0}
    B_0^{\text{frozen}} = {\frac{1}{9}} (c_{11} + c_{22} + c_{33} + 2 c_{12}
    + 2 c_{13} + 2 c_{23})^{\text{frozen}}\\
    = \frac{1}{v} \bigg( C_0 + {\frac{1}{18}} C_1 + {\frac{1}{9}} C_2 \\
    + {\frac{1}{18}} C_3 + {\frac{1}{18}} C_6 + {\frac{1}{18}} C_7 \bigg).
\end{multline}
Using the force constants listed in Table~\ref{tab:force_constants} we find
that there is a roughly 50:50 split between the volume and radial
contributions to $B_0$ in Eq.~(\ref{eq:ptsi_vff_b0}) compared to an
approximate 60:40 split in pure Pt and $\alpha$-Pt$_2$Si.  We already noted
that the volume per atom $v$ is similar in all three materials.  Thus we see
that the smaller value of $C_0$ is partially compensated by an increase in the
radial contribution, yielding a value of $B_0$ which is only slightly smaller
in PtSi than in $\alpha$-Pt$_2$Si, but still approximately a factor of two
larger than in pure Si.  We can now see that the nearly linear relationship
between the bulk modulus and the atomic percent Pt, evident in
Fig.~\ref{fig:elas_const_trends}, has a direct connection with the nature of
the chemical bonding in these materials.  Conversely, the fact that the
volume-conserving elastic constants are similar in magnitude in all four
materials demonstrates that they are less sensitive to the nature of the
bonding.  For example, $c_{44}$ in Pt [Eq.~(\ref{eq:pt_vff_c44})] arises
purely from radial interactions while the split is 90:10, radial to angular,
in Si [Eq.~(\ref{eq:si_vff_c44})].  In the two silicides
[Eqs.~(\ref{eq:pt2si_vff_c44}) and (\ref{eq:ptsi_vff_c44})] the split is
approximately 60:40.  Despite these variations in the split between radial and
angular contributions and variations in the individual force constants
themselves, the volume-conserving elastic constants are all relatively small
and similar in magnitude in all four materials.

\section{Summary}
\label{sec:summary}

We have carried out an extensive study of the chemical bonding and elasticity
of two room-temperature stable platinum silicides, tetragonal
$\alpha$-Pt$_2$Si and orthorhombic PtSi.  We have investigated the trends in
the calculated elastic constants, both the trends within a given material as
well as between materials.  The Cauchy relations, that $c_{12} = c_{66}$,
$c_{13} = c_{55}$, and $c_{23} = c_{44}$, apply to a crystal in which the
interatomic interactions are purely radial.  Real materials deviate from these
relations and we find that in pure Pt as well as the two silicides the
deviations are always positive (left hand side greater than right hand side)
but in Si the deviation is negative.  More generally, we find that in the
metals the elastic constant expressions which correspond to volume-conserving
strains are always smaller than those which correspond to strains which do not
conserve volume.  This also turns out to be true in Si with the exception of
$c_{12}$ which is less than $c_{44}$ (negative deviation from the Cauchy
relation).  However, the difference in magnitudes between volume-conserving
and non-volume-conserving elastic constants is largest on average in Pt and
gets smaller in the progression Pt $\rightarrow$ $\alpha$-Pt$_2$Si
$\rightarrow$ PtSi $\rightarrow$ Si.  In general, the volume-conserving
elastic constants have similar magnitudes in all four materials while the
non-volume-conserving elastic constants follow this same progression.  In
particular, the bulk modulus is found to be a very nearly linear function of
the atomic percentage of Pt.

We have analyzed the valence electronic charge density in order to gain
insight into the nature of the chemical bonding in the silicides.  In the case
of $\alpha$-Pt$_2$Si we find striking evidence of a wide network of covalent
three-center bonds, each involving a single Si atom and two Pt atoms.  Each Si
atom participates in 12 different three-center bonds.  We also find evidence
of two-dimensional metallic Pt (001) sheets which act to interconnect the
network of three-center bonds.  The Pt--Pt bond length in these
two-dimensional sheets is very nearly the same as in pure fcc Pt.  The widely
distributed nature of the bonding in $\alpha$-Pt$_2$Si appears to be closer in
character to the pure metallic bonding in fcc Pt than the covalent two-center
bonds in Si.  The trends in the elastic constants support this interpretation.
PtSi also exhibits evidence of covalent Pt--Si--Pt three-center bonds in
addition to more standard Pt--Si two-center bonds.  Each Si atom participates
in one three-center bond and two two-center bonds with the four Pt neighbors
forming a very distorted tetrahedron.  Two of the six corresponding bond
angles are very nearly equal to the perfect tetrahedral angle but the other
four angles vary from 71$^{\text{o}}$ to 132$^{\text{o}}$.  Qualitatively the
bonding in PtSi appears much more similar to the covalent bonding in pure Si
than the metallic bonding in pure Pt but the trends in the elastic constants
indicate that there are actually elements of both.  The finding of strong
Pt--Si covalent bonding in PtSi is consistent with the experimental study of
Franco \emph{et al.}\cite{Fra01} in which they found spectroscopic evidence
that the influence of the Pt 6$d$ orbitals extends throughout the entire
valence band.

We have constructed valence force field models for the two silicides as well
as pure Pt and pure Si.  These models provide a quantitative basis for
understanding both the trends in the elastic constants and the various
elements of the chemical bonding.  We have included volume-, radial-, and
angular-dependent contributions in the models.  The volume-dependent
contribution, which reflects the presence of metallic bonding, turned out to
be a crucial element of the models.  The presence or absence of this term and
the magnitude of the volume force constant $C_0$ are predominantly responsible
for the observed trend in the non-volume-conserving elastic constants as a
function of Pt concentration.  In addition, the absence of this contribution
in the volume-conserving elastic constants is largely responsible for the fact
that these constants have similar magnitudes in all four materials.  The
variation in the sign and magnitude of the deviations from the Cauchy
relations is a specific example of these more general trends and is once again
due primarily to the variation in the magnitude of $C_0$.  The models also
provide explanations for differences in magnitude between specific elastic
constants for a given material, such as the anomalously large value of
$c_{66}$ in $\alpha$-Pt$_2$Si which we find to be closely connected to the
three-center bonds in this material.

In addition to providing explanations for the trends in the elastic constants,
the magnitudes of the various force constants themselves provide a direct
indication of the nature of the chemical bonding.  The magnitude of the volume
term provides an indication of the relative importance of metallic bonding.
This analysis demonstrated that there is an important element of metallic
bonding in PtSi, despite the lack of direct evidence in the analysis of the
charge density.  This conclusion is required as a result of the specific
values of the elastic constants in this material and would not have been
possible based solely on the qualitative features of the charge density.
Similarly, the magnitudes of the radial and angular force constants are
directly connected to the importance of covalent bonds in the material.  The
trends in these constants confirm the general conclusions made on the basis of
the charge density analysis.  In addition, the conclusion that there are
elements of both metallic and covalent bonding in $\alpha$-Pt$_2$Si as well as
PtSi may be connected to the fact that the heats of formation for the two
silicides are nearly the same.  One general conclusion of this study is that
the elastic constants contain a great deal of information about the nature of
the chemical bonding in a material but since this information is not readily
apparent, an analysis such as the one presented here is necessary in order to
extract the information.  We have attempted to make the case here that an
analysis in terms of valence force field models provides a convenient and
fruitful way to analyze the elastic constants and their connection to the
chemical bonding in a material.

Our purpose in developing the valence force field models described in this
work was to provide a quantitative means for investigating the nature of the
chemical bonding in the platinum silicides in comparison to pure Pt and pure
Si and also to provide a more intuitive understanding of the connection
between the chemical bonding and mechanical properties of these materials.
Nonetheless, we can briefly consider the possibility that these models may
be useful in carrying out future studies of silicide--silicon interfaces where
first principles methods would be vastly more CPU-intensive.  For example,
depending on the growth conditions, the silicide thin film grown on a silicon
substrate can be stabilized in an amorphous phase.  The only hope of treating
such a structure would be to use a more efficient semi-empirical method such
as a valence force field model.  We believe that in general it should be
possible to develop such a model given that our basic formulation includes the
same fundamental elements as in other successful models, such as the
embedded-atom method and Tersoff potentials.  One possible point of concern is
the well-known fact that valence force field models in general tend to
converge very slowly with respect to the number of interaction parameters in
the model.  This issue would certainly need to be explored before any attempt
was made to develop models that could be used in large-scale simulations.

\acknowledgments

This work was performed in part under the auspices of the U.~S. Department of
Energy, Office of Basic Energy Sciences, Division of Materials Science by the
University of California Lawrence Livermore National Laboratory under contract
No.~W--7405--Eng--48.  Partial support was also provided by Deutsche
Forschungsgemeinschaft, SFB 292 ``Multicomponent Layered Systems.''

%\newpage

%\newpage

%\newpage

%\newpage

\widetext

\begin{table}[tbp]
  \caption{Equilibrium theoretical (from Ref.~\protect\onlinecite{Bec01}) and
    experimental lattice constants (in a.u.) and internal structural
    parameters (for PtSi).}
  \begin{tabular}{llddddl}
    Material &        & $a_{0}$ & $b_{0}$ & $c_{0}$ & & Ref.\\
    \hline
    Pt       & theor. &   7.403 &   ---   &   ---   & & \onlinecite{Bec01} \\
             & expt.  &   7.415 &   ---   &   ---   & &
      \onlinecite{LBnsIII/6}\\
    $\alpha$-Pt$_2$Si
             & theor. &   7.407 &   ---   &  11.241 & & \onlinecite{Bec01} \\
             & expt.  &   7.461 &   ---   &  11.268 & & \onlinecite{Ram78} \\
    PtSi     & theor. &  10.583 &  6.774  &  11.195 & & \onlinecite{Bec01} \\
             & expt.  &  10.539 &  6.778  &  11.180 & & \onlinecite{Grae73}\\
    Si       & theor. &  10.22  &   ---   &   ---   & & \onlinecite{Bec01} \\
             & expt.  &  10.26  &   ---   &   ---   & &
      \onlinecite{LBnsIII/29a}\\
    \hline
    PtSi &        & $u_{\text{Pt}}$ & $v_{\text{Pt}}$ & $u_{\text{Si}}$
      & $v_{\text{Si}}$ & Ref.\\
    \hline
         & theor. &      0.9977     &      0.1919     &      0.1782
      &      0.5841     & \onlinecite{Bec01} \\
         & expt.  &      0.9956     &      0.1922     &      0.177
      &      0.583      & \onlinecite{Grae73}\\
  \end{tabular}
  \label{tab:atomic_structure}
\end{table}

\begin{table}[tbp]
  \caption{Elastic constants of Pt and Si.  The first principles calculations,
    described in Ref.~\protect\onlinecite{Bec01}, were carried out at the
    theoretical self-consistent lattice constants of $a_{\text{Pt}} = 7.403$
    a.u. and $a_{\text{Si}} = 10.22$ a.u.  The theoretical value of $c_{44}$
    in parentheses for Si is the ``frozen'' value obtained without allowing
    for internal relaxation.  The bulk modulus is calculated from the elastic
    constants as $B_{0}=\frac{1}{3}(c_{11} + 2 c_{12})$.  Experimental values
    are extrapolated to 0~K.  All values are in units of GPa.}
  \begin{tabular}{l*{4}{r@{}l}}
      & \multicolumn{2}{c}{Pt Theory\protect\cite{Bec01}}
      & \multicolumn{2}{c}{Pt Expt.\protect\cite{Mac65}}
      & \multicolumn{2}{c}{Si Theory\protect\cite{Bec01}}
      & \multicolumn{2}{c}{Si Expt.\protect\cite{LBnsIII/29a}} \\
    \hline
    $c_{11}$ & 346&.8$\pm$0.5           & 358&
             & 163&.45$\pm$0.03         & 165&   \\
    $c_{12}$ & 262&.7$\pm$0.3           & 254&
             &  62&.13$\pm$0.02         &  63&   \\
    $c_{44}$ &  87&.5$\pm$0.3           &  77&
             &  79&.85$\pm$0.02 (108.6) &  79&.1 \\
    $B_{0}$  & 290&.8$\pm$0.3           & 288&.4
             &  95&.90$\pm$0.02         &  97&.0 \\
  \end{tabular}
  \label{tab:pt+si_elastic}
\end{table}

\begin{table}[tbp]
  \caption{First principles elastic constants of $\alpha$-Pt$_2$Si and PtSi
    from Ref.~\protect\onlinecite{Bec01}.  Calculations were performed at the
    theoretical self-consistent lattice constants
    (Table~\ref{tab:atomic_structure}).  ``Frozen'' refers to keeping the
    atoms held fixed at the positions determined solely from the strain tensor
    and, in the case of PtSi, with the internal structural parameters held
    fixed at their theoretical self-consistent values
    (Table~\ref{tab:atomic_structure}).  ``Relaxed'' indicates that a
    relaxation of the atomic positions was carried out, including a relaxation
    of the PtSi internal structural parameters.  Parentheses in the case of
    the relaxed $\alpha$-Pt$_2$Si elastic constants denote values where no
    internal relaxation was necessary because of symmetry constraints (small
    differences with the frozen values come from using a slightly more
    stringent convergence criterion on the energy).  The bulk modulus is
    calculated from the elastic constants as $B_{0}=\frac{1}{9}(2 c_{11} +
    c_{33} + 2 c_{12} + 4 c_{13})$ for $\alpha$-Pt$_2$Si and
    $B_{0}=\frac{1}{9}(c_{11} + c_{22} + c_{33} + 2 c_{12} + 2 c_{13} + 2
    c_{23})$ for PtSi.  No experimental data is available for either material.
    All values are in units of GPa.}
  \begin{tabular}{l*{2}{r@{}l}*{2}{r@{}l}}
      & \multicolumn{2}{c}{$\alpha$-Pt$_2$Si--frozen\protect\cite{Bec01}}
      & \multicolumn{2}{c}{$\alpha$-Pt$_2$Si--relaxed\protect\cite{Bec01}}
      & \multicolumn{2}{c}{PtSi--frozen\protect\cite{Bec01}}
      & \multicolumn{2}{c}{PtSi--relaxed\protect\cite{Bec01}}\\
    \hline
    $c_{11}$ & 347&.2$\pm$1.2 &  332&.4$\pm$0.9  & 327&.5$\pm$1.2
      & 298&.2$\pm$1.2 \\
    $c_{22}$ &    &    ---    &     &    ---     & 313&.8$\pm$0.0
      & 269&.3$\pm$0.8 \\
    $c_{33}$ & 297&.5$\pm$0.5 & (298&.0$\pm$0.4) & 345&.9$\pm$0.1
      & 308&.0$\pm$0.6 \\
    $c_{12}$ & 225&.0$\pm$1.2 &  239&.6$\pm$1.0  & 157&.7$\pm$0.6
      & 156&.4$\pm$0.8 \\
    $c_{13}$ & 169&.3$\pm$0.9 & (169&.4$\pm$0.8) & 162&.9$\pm$0.6
      & 132&.2$\pm$0.7 \\
    $c_{23}$ &    &    ---    &     &    ---     & 153&.4$\pm$0.1
      & 165&.1$\pm$0.6 \\
    $c_{44}$ &  75&.4$\pm$0.3 &   62&.7$\pm$0.5  & 141&.3$\pm$0.3
      & 100&.1$\pm$0.4 \\
    $c_{55}$ &    &    ---    &     &    ---     & 113&.1$\pm$0.1
      & 104&.5$\pm$0.1 \\
    $c_{66}$ & 169&.5$\pm$5.2 & (169&.3$\pm$5.2) &  74&.2$\pm$0.2
      &  66&.3$\pm$0.4 \\
    $B_{0}$  & 235&.4$\pm$0.6 & (235&.5$\pm$0.5) & 215&.0$\pm$0.2
      & 198&.1$\pm$0.3 \\
  \end{tabular}
  \label{tab:silicides_elastic}
\end{table}

%\newpage

\begin{table}[tbp]
  \caption{Force constants of valence force field models
    [Eq.~(\protect\ref{eq:force_field})] for fcc Pt, tetragonal
    $\alpha$-Pt$_2$Si, orthorhombic PtSi, and cubic-diamond phase Si.  The
    $C_0$ force constant represents the volume-dependent interaction, each of
    the remaining $C_i$ is a radial force constant for the $i$th
    nearest-neighbor bond, and $K_{ij}$ is an angular force constant for the
    bond angle between the $i$th and $j$th nearest-neighbor bonds.  All of the
    force constants are in units of eV.}
  \begin{tabular}{ldddd}
              &   Pt  & $\alpha$-Pt$_2$Si &  PtSi &   Si  \\
    \hline
    $C_0$     & 16.54 &       13.88       & 10.38 &  ---  \\
    $C_1$     & 16.10 &       13.58       & 42.58 & 54.06 \\
    $C_2$     &  ---  &       26.03       & 48.26 &  ---  \\
    $C_3$     &  ---  &        8.39       & 10.90 &  ---  \\
    $C_6$     &  ---  &        ---        & 18.32 &  ---  \\
    $C_7$     &  ---  &        ---        &  3.91 &  ---  \\
    $K_{11}$  &  ---  &        5.06       &  ---  &  3.13 \\
    $K_{12}$  &  ---  &        1.87       &  ---  &  ---  \\
    $K_{13}$  &  ---  &        ---        &  1.29 &  ---  \\
    $K_{22}$  &  ---  &        ---        &  7.62 &  ---  \\
    $K_{23}$  &  ---  &        ---        & 15.01 &  ---
  \end{tabular}
  \label{tab:force_constants}
\end{table}

%\printtables
%\printfigures

\begin{references}
\bibitem{rem:notation}
Note that the notation used here differs from that in Ref.~\onlinecite{Har80}
  where $C_0$ refers to the radial force constant and $C_1$ refers to the
  angular force constant.
\end{references}
\end{document}